\documentclass[11pt,fleqn,reqno]{article}
\usepackage[dvips]{graphicx}
\usepackage{epsfig}
\usepackage{sidecap}
\usepackage{amsmath}
\usepackage{bm}
\usepackage{subfigure}
\usepackage{color}
\usepackage[dvips]{graphicx}
\usepackage{epsfig}
\usepackage{amsmath}
\usepackage{bm}
\usepackage{citesort}
\usepackage{subfigure}
\usepackage{color}
\newcommand{\barr}{\bar{r}}

\newcommand{\br}{\mathbf{r}}

\newcommand{\bR}{\mathbf{R}}
\definecolor{gold}{rgb}{0.6,0.3,0.1}

\newcommand{\order}{{\cal O}}
\newcommand{\calJ}{{\cal J}}
\newcommand{\calK}{{\cal K}}
\newcommand{\calH}{{\cal H}}
\newcommand{\calP}{{\cal P}}
\newcommand{\nubar}{\overline{\nu}}
\newcommand{\gind}{g_{ind}}
\newcommand{\gpair}{g_{pair}}
\newcommand{\rhothree}{{\tilde{\rho}}}
\newcommand{\htrue}{h^{true}}
\newcommand{\jtrue}{J^{true}}
\newcommand{\ktrue}{K^{true}}

\begin{document}

\noindent {\large \bf Title:} 

\noindent{\large Pairwise maximum entropy models for studying large biological systems: when they can and when they can't work}\\

\noindent{\large \bf Abbreviated title:} maximum entropy models for biological systems\\

\noindent{\bf Authors:} \\
\noindent Yasser Roudi,\\
Gatsby Computational Neuroscience Unit, UCL, UK\\
\noindent Department of Physiology and Biophysics,\\
Weill Medical College of Cornell University, New York, USA\\

\noindent Sheila Nirenberg,\\
Department of Physiology and Biophysics,\\
Weill Medical College of Cornell University, New York, USA\\

\noindent Peter, E. Latham,\\
Gatsby Computation Neuroscience Unit, UCL, UK\\

\newpage
\noindent {\bf \Large Abstract}

\noindent
One of the most critical problems we face in the study of biological
systems is building accurate statistical descriptions of them. This
problem has been particularly challenging because biological systems
typically contain large numbers of interacting elements, which
precludes the use of standard brute force approaches. Recently,
though, several groups have reported that there may be an alternate
strategy. The reports show that reliable statistical models can be
built without knowledge of all the interactions in a system; instead,
pairwise interactions can suffice. These findings, however, are based
on the analysis of small subsystems. Here we ask whether the
observations will generalize to systems of realistic size, that is,
whether pairwise models will provide reliable descriptions of true
biological systems. Our results show that, in most cases, they will
not. The reason is that there is a crossover in the predictive power
of pairwise models: If the size of the subsystem is below the
crossover point, then the results have no predictive power for large
systems. If the size is above the crossover point, the
results do have predictive power.
This work thus provides a general framework for
determining the extent to which pairwise models can be used to predict
the behavior of whole biological systems.
Applied to neural data, the size of most systems studied so far
is below the crossover point.

\newpage

\section{Introduction}

Many important questions in biology are fundamentally statistical. For
instance, deciphering the neural code requires knowledge of the
probability of observing patterns of activity in response
to stimuli \cite{Rieke97}; determining which features of a protein are important for
correct folding requires knowledge of the probability that a
particular sequence of amino acids folds naturally \cite{Russ05,Socolich05}; and determining
the patterns of foraging of animals and their social and individual
behavior requires knowledge of the distribution of food and species
over both space and time \cite{Oates87,Wrangham87,Eisenberg72}.

Building statistical descriptions of biological systems is, however,
hard. There are several reasons for this: i) biological systems are composed of large numbers of elements, and so
can exhibit a huge number of configurations, in fact, an exponentially
large number, ii) the elements typically interact with
each other, making it impossible to view the system as a collection of
independent entities, and iii) because of technological
considerations, the descriptions of biological systems have to be
built from very little data. For example, with current technology in
neuroscience, we can record simultaneously from only about 100 neurons
out of approximately 100 billion in the human brain. So, not only are
we faced with the problem of estimating probability distributions in
high dimensional spaces, we must make the estimates based on very
little information. 

Despite these apparent difficulties, recent work has suggested that
the situation may be less bleak than it seems. There is evidence that
accurate statistical description of systems can be achieved without having
to examine all possible configurations
\cite{Schneidman05,Shlens06,Tang08,Bethge07,Yu08,Russ05,Socolich05}.
One merely has to measure the probability distribution over pairs of
elements and use those to build the full distribution.
These ``pairwise models'' potentially offer a fundamental
simplification, since the number of pairs is quadratic in the number
of elements, not exponential. However, support for the efficacy of
pairwise models has, necessarily, come from relatively small
subsystems -- small enough that the true probability
distribution could be measured experimentally, allowing direct
comparison of the pairwise distribution to the true one
\cite{Schneidman05,Shlens06,Yu08,Tang08}.
While these studies have provided a
key first step, a critical question remains: will the results from the
analysis of these small subsystems extrapolate to large ones? That is,
if a pairwise model predicts the probability distribution for a subset
of the elements in a system, will it also predict the probability
distribution for the whole system? Here we find that, for a
biologically relevant class of systems, this question can be answered
quantitatively and, importantly, generically -- independent of many of
the details of the biological system under consideration. And the
answer is, generally, ``no.'' In this paper, we explain, both
analytically and with simulations, why this is the case.

\section{Results}

\subsection{The extrapolation problem}
\label{problem:extrapolation}

\begin{SCfigure}
\centering
\includegraphics[height=11cm,width=7cm]
{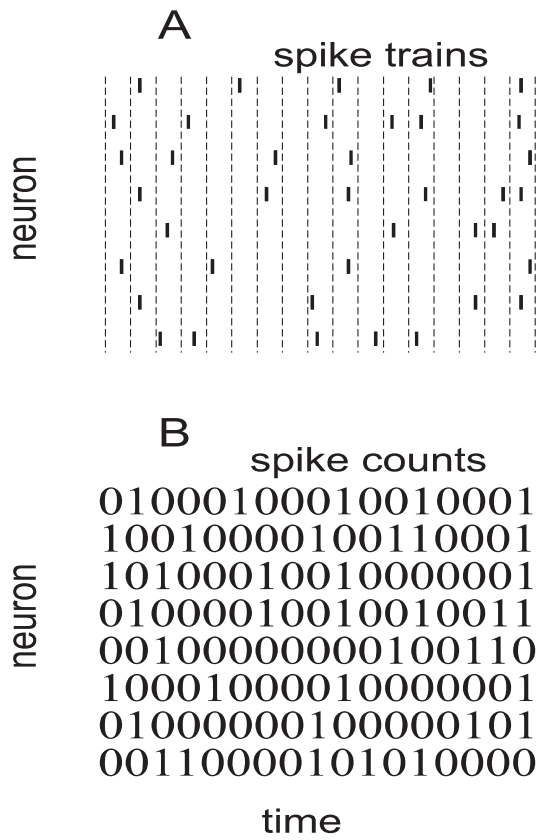}
\caption{Transforming spike trains to spike count.
{\bf A}. Spike rasters. Tick marks indicate spike times; different
rows correspond to different neurons. The horizontal dashed lines are the
bin boundaries.
{\bf B}. Spike count in each bin. In this example the bins are
small enough that there is at most one spike per bin, but this is not necessary -- one
could use bigger bins and have larger spike counts.

\bigskip
\bigskip
\bigskip
\bigskip
\bigskip
\bigskip
}
\label{Fig1}
\end{SCfigure}

To gain intuition into the extrapolation problem, let us consider a
specific example: neuronal spike trains. Figure \ref{Fig1}A shows a
typical spike train for a small population of neurons. Although the
raw spike times provide a complete description, they are not a useful
representation, as they are too high-dimensional. Therefore, we divide time into bins and re-represent the spike train as 0s and 1s: 0 if there is no
spike in a bin; 1 otherwise (Fig.~\ref{Fig1}B)
\cite{Schneidman05,Shlens06,Yu08,Tang08}. For now we assume that the
bins are independent (an assumption whose validity we discuss below,
and in more detail in Sec.~\ref{small-time-bin-wrong}). The problem, then, is
to find $p_{true}(\br) \equiv p_{true}(r_1, r_2, ..., r_N)$ where
$r_i$ is a binary variable indicating no spike ($r_i=0$) or one or
more spikes ($r_i=1$) on neuron $i$. Since this, too, is a high
dimensional problem (though less so than the original spike time
representation), suppose that we instead construct a pairwise
approximation to $p_{true}$, which we denote $p_{pair}$, for a
population of size $N$. (The pairwise model derives its name from the
fact that it has the same mean and pairwise correlations as the true
model.) Our question, then, is: if $p_{pair}$ is close to
$p_{true}$ for small $N$, what can we say about how close the two
distributions are for large $N$?

\begin{SCfigure}
\centering
\includegraphics[height=8cm,width=8cm]
{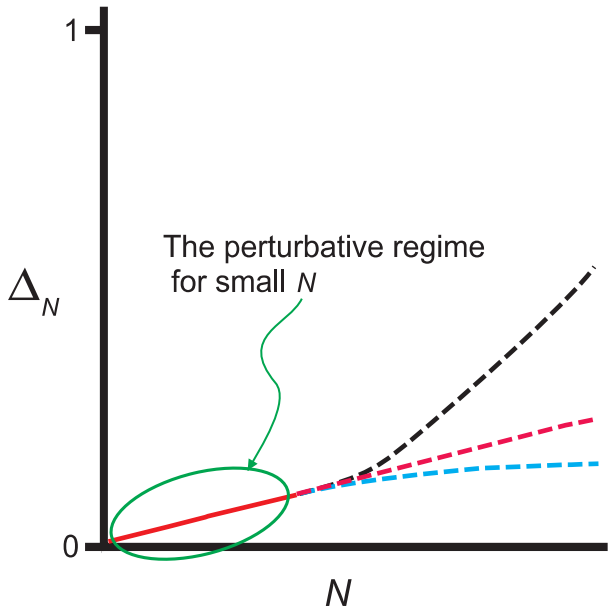}
\caption{Cartoon illustrating the dependence of $\Delta_N$ on
$N$. For small $N$ there is always a
perturbative regime in which $\Delta_N$ increases
linearly with $N$ (solid red line). When $N$ becomes large, on the
other hand, $\Delta_N$
may continue increasing with $N$ (red and black dashed lines) or it may
plateau (cyan dashed line), depending on $p_{true}$. The
observation that $\Delta_N$ increases linearly with $N$ does not,
therefore, provide much, if any
information about the large $N$ behavior.
}
\label{cartoon_DN}
\end{SCfigure}

To answer this question quantitatively, we need a measure of distance.
The measure we use, denoted $\Delta_N$, is defined in Eq.\
(\ref{Delta_N}) below, but all we need to know about it for now is
that if $\Delta_N=0$ then $p_{pair} = p_{true}$, and if $\Delta_N$ is
near one then $p_{pair}$ is far from $p_{true}$. In terms of
$\Delta_N$, our main results are as follows: first, for small $N$, in
what we call the perturbative regime, $\Delta_N$ is proportional to
$N-2$. In other words, as the population size increases, the pairwise
model becomes a worse and worse approximation to the true
distribution. Second, this behavior is entirely generic: for small
$N$, $\Delta_N$ increases linearly, no matter what the true
distribution is. This is illustrated schematically in
Fig.~\ref{cartoon_DN}, which shows the generic behavior of $\Delta_N$.
The solid red part of the curve is the perturbative regime, where
$\Delta_N$ is a linearly increasing function of $N$; the dashed curves
show possible behavior beyond the perturbative regime.

These results have an important corollary: if one does an
experiment and finds that $\Delta_N$ is increasing linearly with $N$, then
one has no information at all about the true distribution. The flip
side of this is more encouraging: if one can measure the true
distribution for sufficiently large $N$ that $\Delta_N$ saturates, as
in the dashed blue line in Fig.~\ref{cartoon_DN}, then one can have
some confidence that extrapolation to large $N$ {\em is} meaningful.
The implications for the interpretation of experiments is, then, that
extrapolation to large $N$ is valid only if one can analyze data past
the perturbative regime.

Under what conditions is a subsystem in the perturbative regime? The
answer turns out to be simple: the size of the system, $N$, times the
average probability of observing a spike in a bin, must be small
compared to $1$. For example, if the average probability is $1/100$,
then a system will be in the perturbative regime if the number of
neurons is small compared to $100$. This last observation would seem
to be good news for studies in which spikes are binned across time and
temporal correlations are ignored. For such binned spike trains, the
probability of a spike can be made arbitrarily small by simply
shrinking the time bins, and so the size of the population for which
the pairwise model appears good can be made arbitrarily large. The
problem with this, though, is that temporal correlations can be
ignored only when time bins are large compared to the autocorrelation
time. This leads to a kind of catch-22: pairwise models are guaranteed
to work well (in the sense that they describe spike trains in which
temporal correlations are ignored) if one uses small time bins, but
small time bins is the one regime where ignoring temporal
correlations is not a valid approximation.

In the next several sections we quantify the qualitative picture
presented above: we write down an
explicit expression for $\Delta_N$, explain why it increases linearly
with $N$, when $N$ is small, and provide additional tests, besides
assessing the linearity of $\Delta_N$, to determine whether or not one
is in the perturbative regime. 

\subsection{A measure of goodness of fit}

A natural measure of the distance between $p_{pair}$ and $p_{true}$ is
the Kullback-Leibler (KL) divergence \cite{Kullback51}, denoted
$D_{KL}(p_{true}||p_{pair})$ and defined as

\begin{equation}
D_{KL}(p_{true}||p_{pair})=\sum_{\br}  p_{true} (\br) \log_2
{p_{true}(\br) \over p_{pair}(\br)}
\, .
\label{KL-def}
\end{equation}

\noindent
The KL divergence is zero if the two distributions are equal; otherwise
it is nonzero.

Although the KL divergence is a very natural measure, it is not easy
to interpret (except, of course, when it is exactly zero). That is
because a nonzero KL divergence tells us is that $p_{pair} \ne
p_{true}$, but it does not give us any real handle on how much we
benefit by including the pairwise correlations in our approximation. To 
make sense of the KL divergence, then, we need something to compare 
it to. A reasonable reference quantity, used by a number of authors
\cite{Schneidman05,Shlens06,Tang08}, is the KL divergence between the
true distribution and the independent one, the latter denoted
$p_{ind}$. The independent distribution, as its name suggests, is a
distribution in which the variables are taken to be independent,
\begin{equation} p_{ind}(r_1,\dots,r_N)= \prod_i p_i(r_i) \, ,
\label{ind} \end{equation}

\noindent
where $p_i(r_i)$ is the distribution of the response of the $i^{\rm
th}$ neuron, $r_i$. With this choice for a comparison, we define our
measure of goodness of fit as
\begin{equation}
\Delta_N= \frac{D_{KL}(p_{true}||p_{pair})}{D_{KL}(p_{true}||p_{ind})}.
\label{DeltaN-def}
\end{equation}

\noindent
Note that the denominator in this expression,
$D_{KL}(p_{true}||p_{ind})$, is usually referred to as
the multi-information \cite{Friedman01,Schneidman05,Slonim06}.

The quantity $\Delta_N$, which we introduced in the previous section,
lies between 0 and 1, and measures how well a pairwise model does
relative to an independent model. If it is 0, the pairwise model is
equal to the true model ($p_{pair}(\br)=p_{true}(\br)$); if it is near
1, the pairwise model offers little improvement over the independent
model; and if it is exactly 1, the pairwise model
is equal to the independent model
($p_{pair}(\br)=p_{ind}(\br)$), and so offers no improvement.
(Our assertion that $\Delta_N$ cannot
exceed 1 assumes that the pairwise model cannot be worse than the
independent one, something that is reasonable in practice but not
guaranteed in general.)

How do we attach physical meaning to the two divergences
$D_{KL}(p_{true}||p_{pair})$ and $D_{KL}(p_{true}||p_{ind})$?
For the latter, we use the fact that, as is easy to show,
\begin{equation}
D_{KL}(p_{true}||p_{ind})=S_{ind}-S_{true},
\label{KL-Ent-ind}
\end{equation}

\noindent
where $S_{ind}$ and $S_{true}$ are the entropies
\cite{Shannon49,Cover91} of $p_{ind}$ and
$p_{true}$, respectively, defined, as usual,
to be $S[p] = -\sum_{\br} p(\br)
\log_2 p(\br)$.
For the former, we use the definition of the KL divergence to write
\begin{equation}
D_{KL}(p_{true}||p_{pair})
=
- \sum_{\br}  p_{true} (\br) \log_2 (p_{pair}(\br))
- S_{true}
\equiv
\tilde{S}_{pair}-S_{true}
\, .
\label{KL-Ent-maxent}
\end{equation}

\noindent
The quantity $\tilde{S}_{pair}$ has the flavor of an entropy, although
it is a true entropy only
when $p_{pair}$ is maximum entropy as well as pairwise (the
maximum entropy pairwise model, or maximum entropy model for short;
see Eq.~\eqref{maxent}). For other pairwise distributions,
all we need to know is that $\tilde{S}_{pair}$ lies between $S_{true}$
and $S_{ind}$, something that is guaranteed by our assumption that the
pairwise model is no worse than the independent model.

\begin{SCfigure}[][t!!!]
\epsfig{file=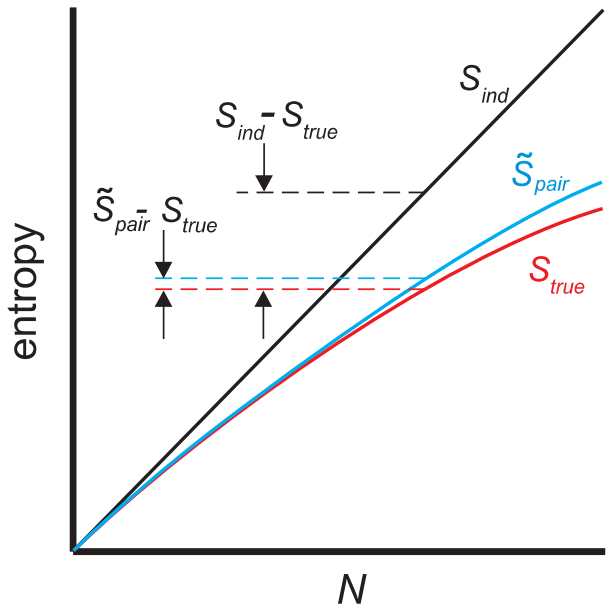,height=8cm,width=8cm}
\caption{Schematic plot of
$S_{ind}$ (black line), $\tilde{S}_{pair}$ (cyan line) and $S_{true}$
(red line). The better the pairwise model, the closer
$\tilde{S}_{pair}$ is to $S_{true}$. This is reflected in the cost
function $\Delta_N$, which is the distance between the red
and cyan lines divided by the distance between the red and black
lines.
\bigskip
\bigskip
}
\label{entropies}
\end{SCfigure}

What Eqs.~\eqref{KL-Ent-ind}
and \eqref{KL-Ent-maxent} tell us is that
$\Delta_N$ (Eq.~\eqref{DeltaN-def}) is the ratio of the amount 
of ``entropy'' {\em not} explained by the pairwise 
model to the amount of entropy {\em not} explained by the 
independent model. A plot illustrating the relationship 
between $\Delta_N$, the two entropies $S_{ind}$ and $S_{true}$, and
the entropy-like quantity $\tilde{S}_{pair}$,
is shown in Fig.~\ref{entropies}.

\subsection{$\Delta_N$ in the perturbative regime}
\label{N-dependence}

The extrapolation problem discussed above is the problem of
determining $\Delta_N$ in the large $N$ limit. This is hard to do
in general, but, as we show in
Methods, Sec.~\ref{perturbative_expansion},
there is a perturbative regime in which it is possible.
The small parameter that defines this regime is the average number of
spikes per bin; this is written quantitatively as $N \nubar \delta t$
where $\delta t$ is the bin size and
$\nubar$ the average firing rate. Letting
and $\nu_i$ the firing rate of neuron $i$, the latter quantity is
given by
\begin{equation}
\label{nubar}
\nubar \equiv {1 \over N} \sum_i \nu_i
\, .
\end{equation}

\noindent
Note that our small parameter depends on $N$, which means that
for $\nubar$ and $\delta t$ fixed, there is a maximum population
size we can access perturbatively. We return to this point below, as
it has major implications for experimental studies.

The first step in the perturbation expansion is to compute the
two quantities that make up $\Delta_N$: $D_{KL}(p_{true}||p_{ind})$
and $D_{KL}(p_{true}||p_{pair})$. As we show in
Methods, Sec.~\ref{perturbative_expansion}, these are given by
\begin{subequations}
\begin{align}
D_{KL}(p_{true}||p_{ind})&=\gind N(N-1)\ (\nubar \delta t)^2 +
\order \left((N \nubar \delta t)^3 \right) \label{KLs_Na}
\\
D_{KL}(p_{true}||p_{pair})&=\gpair N(N-1)(N-2)\ (\nubar \delta t)^3 +
\order \left((N \nubar \delta t)^4 \right) \label{KLs_Nb}
\, .
\end{align}
\label{KLs_N}
\end{subequations}

\noindent
The exact forms of the prefactors $\gind$ and $\gpair$ are given in
Eqs.\ (\ref{gind}) and (\ref{gpair}). The details, however, are not so
important; the important
things to know about them is that they are
independent of $N$ and  $\nubar \delta t$, and they depend on the low
order statistics of the spike trains: $\gind$ depends on the second
order normalized correlations function, and $\gpair$ depends on 
the second and third order normalized correlations function, as 
defined below in Eq.~\eqref{rho_norm}.
The $N$-dependence in the first term on
the right hand side of Eq.~\eqref{KLs_Na} has
been noted previously \cite{Schneidman05}, although the authors did
not compute the prefactor, $\gind$.

Inserting Eq.~\eqref{KLs_N} into Eq.~\eqref{DeltaN-def} (into the
definition of $\Delta_N$), we arrive at our main result,
\begin{equation}
\Delta_N=\frac{\gpair}{\gind} (N-2)\ \nubar \delta t +
\order \left((N \nubar \delta t)^2 \right)
\, .
\label{Delta_N}
\end{equation}

\noindent
This expression tells us how $\Delta_N$ scales with $N$ in the
perturbative regime --
the regime in which $N \nubar \delta t \ll 1$.
The key observation about this scaling is that it is
independent of the details of the true distribution, $p_{true}$. This
has a very important consequence, one that has major implications for
experimental data: if one does an experiment and finds that that
$\Delta_N$ is proportional to $N-2$, then the system is in the
perturbative regime, and one does not know whether
$p_{pair}$ will remain close to $p_{true}$ as $N$ increases.
What this means in practical terms is that if one wants to know
whether a particular pairwise
model is a good one for large systems,
it is necessary to consider values of $N$ that are significantly
greater than $N_c$, where
\begin{equation}
N_{c} \equiv \frac{1}{\nubar \delta t}
\, .
\label{N-c}
\end{equation}
We interpret $N_c$ as the value at which there is a crossover
in the behavior of the pairwise model. 
Specifically, if $N \ll N_c$, the system is in the perturbative
regime and the pairwise model is not informative
about the large $N$ behavior, whereas if $N \gg N_c$, the 
system is in a regime in which it may be
possible to make inferences about the behavior of the full system.

\subsection{The dangers of extrapolation}
\label{dangers:extrapolation}

Although the behavior of $\Delta_N$ in the perturbative regime does
not tell us much about its behavior at large $N$, it
is possible that other quantities that can be calculated in
the perturbative regime, $\gind$, $\gpair$,
and $S_{ind}$ (the last one exactly), are informative,
as others have suggested \cite{Schneidman05}. Here we
show that they are also uninformative.

\begin{figure}
\epsfig{file=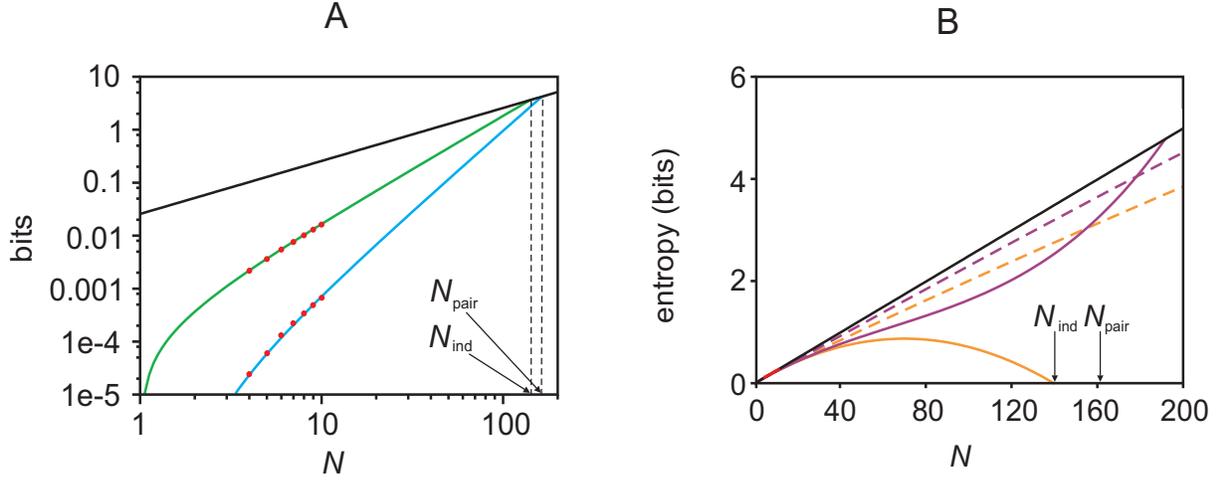,height=6.5cm,width=16cm}
\caption{Cartoon showing the extrapolation of entropy and KL
divergences, and illustrating why the two natural quantities
derived from it, $N_{ind}$ and $N_{pair}$, occur beyond the point at which the
extrapolation is meaningful.
{\bf A.} Extrapolations on a log-log scale.
Black: $S_{ind}$ versus $N$; green: extrapolation of
$D_{KL}(p_{true}||p_{ind})$;
cyan: extrapolation of $D_{KL}(p_{true}||p_{maxent})$.
The red points are the data.
The points $N_{ind}$ and $N_{pair}$ label the
intersections of the two extrapolations with the independent entropy,
$S_{ind}$.
{\bf B.} Extrapolation of the entropies rather than the KL
divergences, plotted on a linear-linear scale.
The data, again shown in red, is barely visible in the lower left hand
corner. Black: $S_{ind}$ versus $N$; solid maroon: extrapolation of
$\tilde{S}_{pair}$;
solid orange: extrapolation of $S_{true}$.
The dashed maroon and orange lines are the extrapolations
of the true pairwise ``entropy'' and true entropy, respectively.
\bigskip
\bigskip
}
\label{extrapolation}
\end{figure}

The easiest way to relate the perturbative regime to the large $N$
regime is to extrapolate Eqs.~(\ref{KLs_N}a) and (\ref{KLs_N}b), and
ask what their large $N$ behavior tells us. Generic versions of
these extrapolations, plotted on a log-log scale, are shown in
Fig.~\ref{extrapolation}A, along with a plot of the independent
entropy, $S_{ind}$ (which is necessarily linear in $N$; see
Sec.~\ref{s_asymptotic}). The first thing we notice about the
extrapolations is that they do not, technically, have a large $N$
behavior: one terminates at the point labeled $N_{ind}$, which is
where $D_{KL}(p_{true}||p_{ind}) = S_{ind}$ (and thus $S_{true} = 0$;
continuing the extrapolation implies negative true entropy);
the other at the point labeled $N_{pair}$, which is where
$D_{KL}(p_{true}||p_{pair})=S_{ind}$ (and thus $S_{true} \le 0$, since
$\tilde{S}_{pair} \le S_{ind}$).

Despite the fact that the extrapolations end abruptly, they still
might provide information about the large $N$ regime. For example,
$N_{pair}$ and/or $N_{ind}$ might be values of $N$ at which something
interesting happens. To see if this is the case, in
Fig.~\ref{extrapolation}B we plot the naive extrapolations of
$\tilde{S}_{pair}$ and $S_{true}$, as given by Eq.~\eqref{KLs_N}, on a
linear-linear plot, along with $S_{ind}$ (solid lines). This plot
contains no new information compared to Fig.~\ref{extrapolation}A, but
it does elucidate the meaning of the extrapolations. Perhaps its most
striking feature is that the naive extrapolation of $S_{true}$ has a
decreasing portion. As is easy to show mathematically, this cannot
happen (intuitively, that is because observing one additional neuron
cannot decrease the entropy of previously observed neurons). Thus,
$N_{ind}$, which occurs well beyond the point where the naive
extrapolation of $S_{true}$ is decreasing, has essentially no meaning,
something that has been pointed out previously
by Bethge et al. \cite{Bethge07}. The other potentially important value of $N$ is $N_{pair}$.
This, though, suffers from similar problems: either
$N_{pair}>N_{ind}$, in which case the entropy is
negative, or it crosses the green curve in Fig.~\ref{extrapolation}A
from below, meaning $\Delta_N > 1$. Either way, it also doesn't have much
meaning.

How do the naively extrapolated entropies -- the solid lines in
Fig.~\ref{extrapolation}B -- compare to the true entropies? To answer this, in
Fig.~\ref{extrapolation}B we show the true behavior of $S_{true}$ and
$\tilde{S}_{pair}$ versus $N$ (dashed lines). Note that $S_{true}$ is
asymptotically linear in $N$, even though the neurons are correlated,
a fact that forces $\tilde{S}_{pair}$ to be linear in $N$, as it is
sandwiched between $S_{true}$ and $S_{ind}$. (The asymptotically
linear behavior of $S_{true}$ is typical, even in highly correlated systems.
Although this is not always appreciated, it is easy to show; see
Sec.~\ref{s_asymptotic}.) Comparing the dashed and solid lines, we see
that the naively extrapolated and true entropies, and thus the naively
extrapolated and true values of $\Delta_N$, have extremely different
behavior. This further suggests that there is very little connection
between the perturbative and large $N$ regimes.

These observations can be summarized by noting that $\gind$ and
$\gpair$ depend only on correlation coefficients up to third order
(see Eqs.~\eqref{gind} and \eqref{gpair}),
whereas the large $N$ behavior depends on
correlations at all orders. Thus, since $\gind$ and $\gpair$ tell us very
little, if anything, about higher order correlations, it is not
surprising that they tell us very little about the behavior of
$\Delta_N$ in the large $N$ limit.

\subsection{Numerical simulations}
\label{numerical_simulations}

\begin{figure}
\centerline{
\epsfig{file=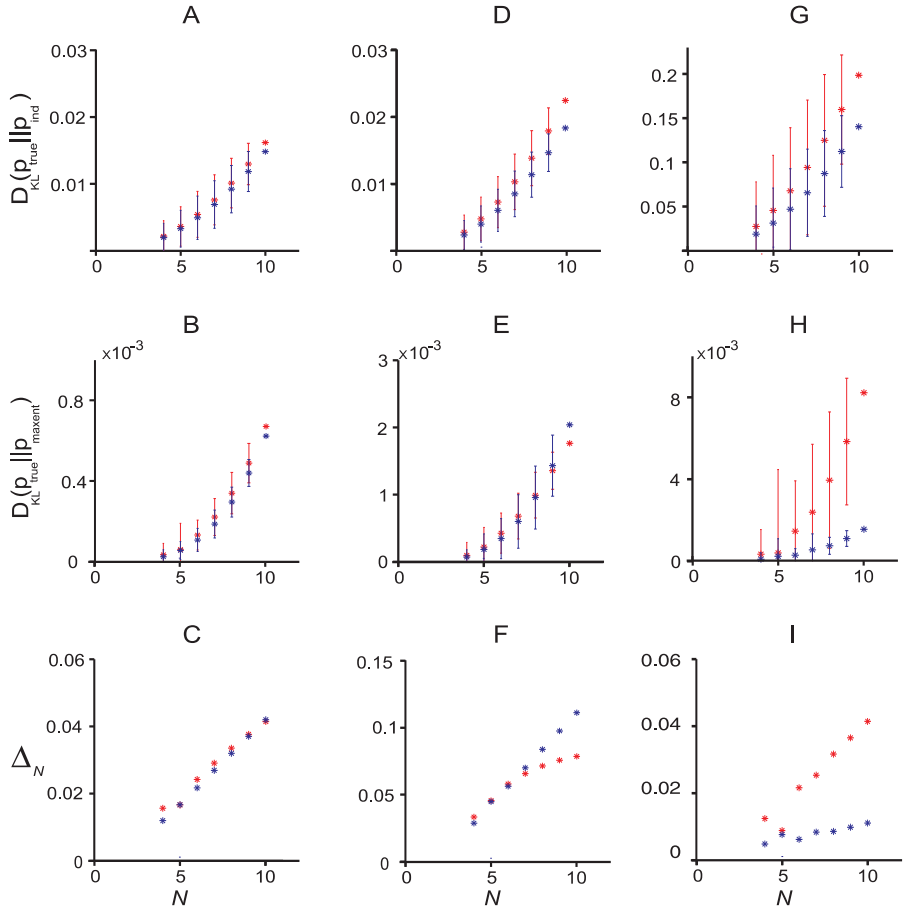,height=12cm,width=12cm}
}
\caption{The $N$ dependence of the KL divergences and the goodness of
fit, $\Delta_N$. Data was generated from a third order model, as
explained in Sec.~\ref{Generating_syn_data} (Methods), and
fit to maximum entropy pairwise model and independent
models. All data points correspond to averages over
marginalizations of the true distribution
(see text for details).
The red points were computed directly from the model fits,
Eqs.~\eqref{KL-def}, \eqref{DeltaN-def} and \eqref{KL-Ent-ind};
the blue points are predictions of the perturbative
expansions,
Eqs.~\eqref{KLs_N} and \eqref{Delta_N}.
The three columns correspond to $\nubar \delta t =$ 0.024, 0.029, and
0.037, from left to right.
{\bf A}, {\bf B}, {\bf C} ($\nubar \delta t = 0.024$).
Predictions from the perturbative expansion are in good agreement with
the measurements up to $N=10$,
indicating that the data is in the perturbative regime.
{\bf D}, {\bf E}, {\bf F} ($\nubar \delta t = 0.029$).
Predictions from the perturbative expansion are in good agreement with
the measurements up to $N=7$, indicating that the data is only
partially in the perturbative regime.
{\bf G}, {\bf H}, {\bf I} ($\nubar \delta t = 0.037$).
Predictions from the perturbative expansion are not in good agreement with
the measurements, even for small $N$, indicating that the data is
outside the perturbative regime.
}
\label{KL_Delta}
\end{figure}

To check that our perturbation expansions, Eqs.~\eqref{KLs_N}
and \eqref{Delta_N}, are correct, and to
investigate the regime in which they are valid, we performed numerical
simulations. We generated, from synthetic data, a set of true
distributions, computed $D_{KL}(p_{true}||p_{ind})$,
$D_{KL}(p_{true}||p_{pair})$, and $\Delta_N$ numerically for each of
them, and compared to the values predicted by Eqs.~\eqref{KLs_N} and
\eqref{Delta_N}. The results are shown in Fig.~\ref{KL_Delta}. Before
discussing that figure, though, we explain our procedure for
constructing true distributions.

The set of true distributions we used were generated from a
third order model (so named because it includes up to third
order interactions). This model has the form
\begin{equation}
p_{true}(r_1, \dots, r_{N^*}) =
\frac{1}{Z_{true}}
\exp\left[
\sum_i \htrue_i r_i +
\sum_{i < j} \jtrue_{ij} r_i r_j +
\sum_{i < j < k} \ktrue_{ijk} r_i r_j r_k
\right]
\label{ptrue}
\end{equation}

\noindent where $Z_{true}$ is a normalization constant, chosen to ensure the the
probability distribution sums to 1, and the sums over $i$, $j$ and $k$
run from 1 to $N^*$. The parameters $\htrue_i, \jtrue_{ij}$ and
$\ktrue_{ijk}$ were chosen by sampling from distributions (see
Methods, Sec.~\ref{Generating_syn_data}), which allowed us
to generate many different true distributions.

For a particular simulation (corresponding to a column in
Fig.~\ref{KL_Delta}), we generated a true distribution with $N^*=15$,
randomly chose 5 neurons, and marginalized over them. This gave us a
10-neuron true distribution. True distributions with $N < 10$ were
constructed by marginalizing over additional neurons within our
10-neuron population. To achieve a
representative sample, we considered all possible marginalizations
(of which there are $10$ choose $N$, or $10!/[N!(10-N)!]$).
The results in Fig.~\ref{KL_Delta} are averages over these
marginalizations.

For neural data, the most commonly
used pairwise model is the maximum entropy model. Therefore, we use that
one here. To emphasize the maximum entropy nature of this model,
we replace the label ``{\em pair}'' that we
have been using so far with ``{\em maxent}.''
The maximum entropy distribution has the form

\begin{equation}
p_{maxent}(\br) =
\frac{1}{Z}
\exp\left[
\sum_i h_i r_i + \sum_{i < j} J_{ij} r_i r_j
\right]
\, .
\label{maxent}
\end{equation}

Fitting this distribution requires that we choose the $h_i$
and $J_{ij}$ so that the first and second moments match those of the
true distribution. Quantitatively, these conditions are

\begin{subequations}
\begin{align}
\langle r_i \rangle_{maxent} & = \langle r_i \rangle_{true}
\\
\langle r_i r_j \rangle_{maxent} & = \langle r_i r_j \rangle_{true}
\end{align}
\label{pair-fit}
\end{subequations}

\noindent
where the angle brackets, $\langle \dots \rangle_{maxent}$ and
$\langle \dots \rangle_{true}$, represent averages with respect to
$p_{maxent}$ and $p_{true}$, respectively.
Once we have $h_i$ and $J_{ij}$ that satisfy Eq.~\eqref{pair-fit}, we calculate the KL
divergences, Eqs. (\ref{KLs_N}), and use those to compute $\Delta_N$.

The results are shown in Fig.~\ref{KL_Delta}. The rows correspond to our three
quantities of interest: $D_{KL}(p_{true}||p_{ind})$,
$D_{KL}(p_{true}||p_{pair})$, and $\Delta_N$ (top to bottom). The
columns correspond to different values of $\nubar \delta t$, with the
smallest $\nubar \delta t$ on the left and the largest on the right.
Red circles are the actual values of these quantities; blue ones are
the predictions from Eqs.~\eqref{KLs_N} and \eqref{Delta_N}. 

As suggested by our perturbation analysis, the smaller the value of
$\nubar \delta t$, the better the agreement between computed and
predicted values. Our simulations corroborate this: the left column of
Fig.~\ref{KL_Delta}
has $\nubar \delta t = 0.024$, and agreement is almost perfect out
to $N=10$; the middle column has $\nubar \delta t = 0.029$, and
agreement is almost perfect out to $N=7$; and the right column
has $\nubar \delta t = 0.037$, and agreement is not good for any
value of $N$.

These results validate the perturbation expansions in
Eqs.~\eqref{KLs_N} and \eqref{Delta_N}, and show that those expansions
provide sensible predictions -- at least for some parameters. They
also suggest a natural way to assess the significance of one's data:
plot $D_{KL}(p_{true}||p_{ind})$, $D_{KL}(p_{true}||p_{pair})$, and
$\Delta_N$ versus $N$, and look for agreement with the predictions of
the perturbation expansion. If agreement is good, as in the left
column of Fig.~\ref{KL_Delta}, then one is in the perturbative regime, and one
knows very little about the true distribution. If, on the other hand,
agreement is bad, as in the right column, then one is out of the
perturbative regime, and there is hope of extracting meaningful
information about the relationship between the true and pairwise
models.

That said, the qualifier ``at least for some parameters'' is an
important one. This is because the perturbation
expansion is essentially an expansion that depends on
the normalized correlation coefficients, and there is an
underlying assumption that they don't
exhibit pathological behavior. The $k^{\rm th}$ normalized
correlation coefficient for the distribution $p$,
denoted $\rho^p_{i_1 i_2 \dots i_k}$, is written

\begin{equation}
\rho^p_{i_1 i_2 \dots i_k} \equiv
\frac{
\langle
(\langle r_{i_1}  - \langle r_{i_1} \rangle_p)
(\langle r_{i_2}  - \langle r_{i_2} \rangle_p)
\dots
(\langle r_{i_k}  - \langle r_{i_k} \rangle_p)
\rangle_p
}
{
\langle r_{i_1} \rangle_p
\langle r_{i_2} \rangle_p
\dots
\langle r_{i_k} \rangle_p
}
\, .
\label{rho_norm}
\end{equation}

\noindent
A potentially problematic feature
of the correlation coefficients is that the denominator is
a product over mean activities. If the mean activities are
small, the
denominator can become very small, leading to very large correlation
coefficients. Although our perturbation expansion is always
valid for sufficiently small time bins (because the correlation
coefficients eventually becomes independent of bin size; see Methods,
Sec.~\ref{Bin-size}),  ``sufficiently small'' can depend in detail on the
parameters. For instance, at the maximum population size tested 
($N=10$) and for the true distributions that had
$\nubar \delta t < 0.03$,  the absolute 
error of the prediction had a median of approximately $16\%$. However,
about 11\% of the runs had errors larger than $60\%$.
Thus, the exact size of the small parameter at which the
perturbative expansion breaks down can depend on the details of the
true distribution.

\subsection{Local fields and pairwise couplings have a simple
dependence on firing rates and correlation coefficients
in the perturbative regime}

\label{fields-Js}

Estimation of the KL divergences and $\Delta_N$ from real data can be
hard, in the sense that it takes a large amount of data for them
to converge to their true values. We therefore provide a
second set of relationships that can be used to determine whether or
not a particular data set is in the perturbative regime. These
relationships are between the parameters of the maximum entropy model,
the $h_i$ and $J_{ij}$, and the mean activity and normalized second
order correlation coefficient (the latter defined in
Eq.~\eqref{rho_norm}).

Since the quantity $\nubar \delta t$ plays a central role in our
analysis, we replace it with a single parameter, which we denote
$\delta$,
\begin{equation}
\delta \equiv \nubar \delta t
\, .
\label{delta}
\end{equation}
\noindent
In terms of this parameter,
we find, (using the same perturbative approach that led us to Eqs.\
(\ref{KLs_N}) and (\ref{Delta_N}); see Sec.~\ref{local-fields}), that
\begin{subequations}
\begin{align}
h_i &=
\log \left[ \langle r_i \rangle^{-1} -1\right]+ \order(N\delta) 
\\
J_{ij} &=
\log \left[ 1+\rho_{ij} \right]+ \order(N \delta)
\end{align}
\label{hJ}
\end{subequations}

\noindent
where $\rho_{ij}$, the normalized second order
correlation coefficient, is defined in Eq.~\eqref{rho_norm} with
$k=2$; it is given explicitly by
\begin{equation}
\rho_{ij}=\frac{\langle r_i r_j \rangle-\langle r_i \rangle_ \langle
r_j \rangle}{\langle r_i \rangle_ \langle r_j \rangle}
\, .
\label{norm-corr}
\end{equation}

\noindent
(We don't need a superscript on $\rho$ or a subscript on the
angle brackets because the first and second moments are
the same under the true and pairwise distributions.)

\begin{figure}
\epsfig{file=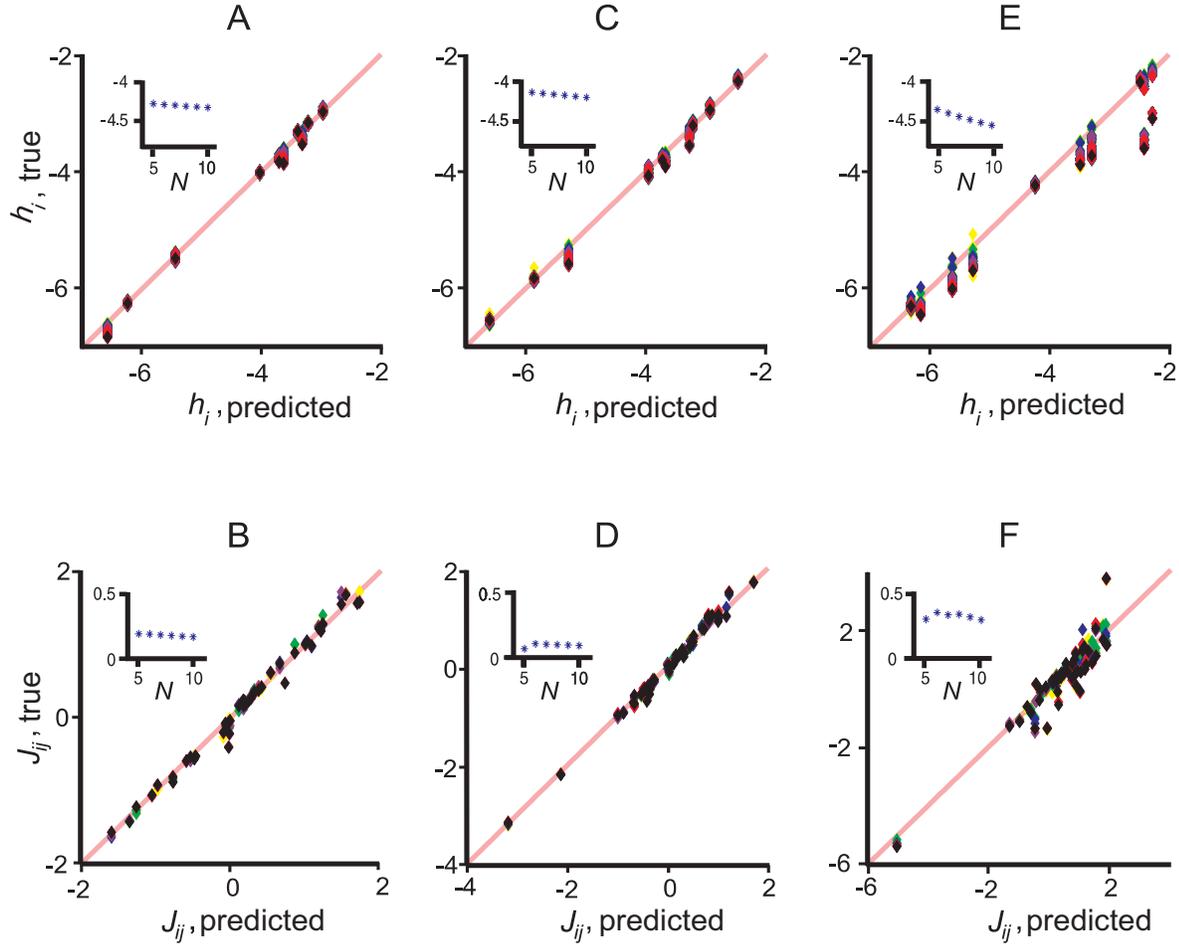,height=12.5cm,width=15.5cm}
\caption{Comparison of the true local fields ($h_i$, top row) and pairwise
interactions ($J_{ij}$, bottom row), to
the predictions from the perturbation expansion, Eq.~\eqref{hJ}.
Values of $N$ ranging from 5 to 10 are shown, with different
colors corresponding to different $N$s.
For each value of $N$, fits are shown for 45 realization of the true
distribution. Insets show the $N$-dependence of the mean
local fields (top) and mean pairwise interactions (bottom).
The three columns correspond exactly to the columns in Fig.~\ref{KL_Delta}.
{\bf A}, {\bf B} ($\nubar \delta t = 0.024$).
There is a very good match between the theoretical and fit values
of both local fields and pairwise interactions.
{\bf C}, {\bf D} ($\nubar \delta t = 0.029$).
Even though $\nubar \delta t$ has increased, the match is still good.
{\bf E}, {\bf F} ($\nubar \delta t = 0.037$).
The predicted and fit local fields and pairwise interactions do not match
as well as the cases shown in A,B, C and D. There is also now a
strong $N$-dependence in the mean local
fields, and a somewhat weaker dependence in the pairwise interactions.
This indicates that the perturbative expansion is breaking down.
}
\label{h_J_N}
\end{figure}

Equation \eqref{hJ} tells us that the $N$-dependence of the
$h_i$ and $J_{ij}$, the local fields and pairwise couplings, are very
weak. In Fig.~\ref{h_J_N} we confirm these theoretical predictions
through numerical simulations.

As an aside, we should point out
that the $N$-dependence is a function of the
variables used to represent the firing patterns. Here we use 0 for silence and 1
for firing, but another, possibly more common,
representation, derived from the Ising model
and used in a number of studies \cite{Schneidman05,Tang08,Yu08}, is to use $-1$ for silence and $+1$ for firing. This
amounts to making the change of variables $s_i = 2r_i-1$. In terms of
$s_i$, the maximum entropy model has the form
$p(\br) \sim \exp \left[ \sum_i h^{ising}_i s_i + \sum_{i < j}
J^{ising}_{ij} s_i s_j \right]$ where 
$h_i^{ising}$ and $J_{ij}^{ising}$ are given by
\begin{subequations}
\begin{align}
h_i^{ising} &= \frac{h_i}{2} +
\sum_{j \ne i} \frac{J_{ij}}{4}
\\
J_{ij}^{ising} &= \frac{J_{ij}}{4}
\, .
\end{align}
\label{ising}
\end{subequations}

\noindent
The second term on the right hand side of Eq.~(\ref{ising}a) is
proportional to $N-1$, which means the local fields in the Ising
representation acquire a linear $N$-dependence that was not present in
our 0/1 representation. The two
studies that reported the $N$-dependence of the local fields
\cite{Schneidman05,Tang08} used this
representation, and, as predicted, their local fields
had a component that was linear in $N$.

\begin{figure}
\centering
\epsfig{file=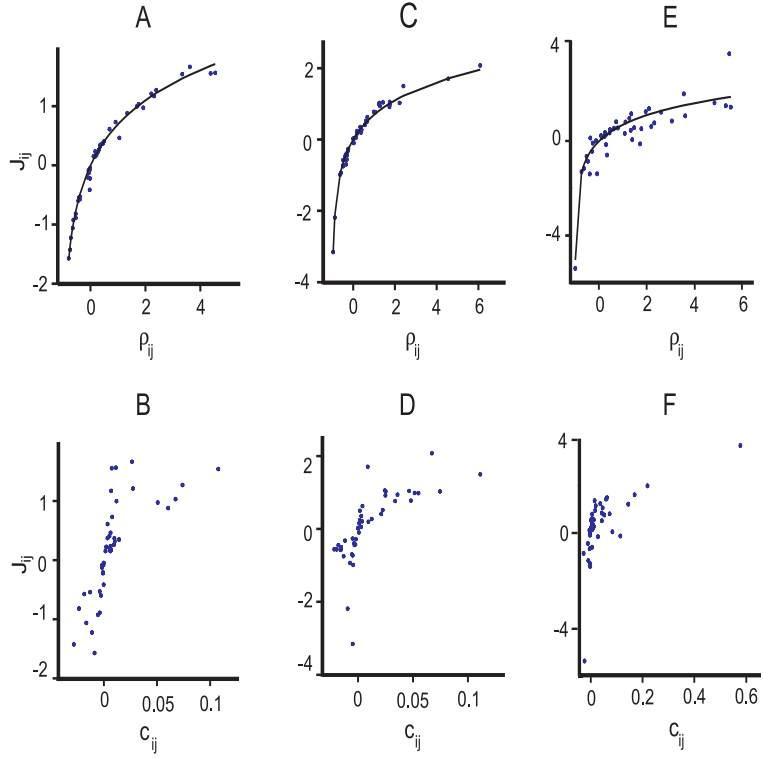,height=10cm,width=10cm}
\caption{Pairwise couplings versus pairwise correlations,
showing that there is a simple relation between
$J_{ij}$ and $\rho_{ij}$ but not between $J_{ij}$ and $c_{ij}$. 
Top row: $J_{ij}$ versus the normalized coefficients,
$\rho_{ij}$ (blue points), along with predicted relationship, via
Eq.~(\ref{hJ}b) (black line).
Bottom row: $J_{ij}$ versus the Pearson correlation
coefficients, $c_{ij}$, Eq.~\eqref{c-ij} (blue points).
The three columns correspond exactly to the columns in
Fig.~\ref{KL_Delta}, for which $\nubar \delta t$ = 0.024, 0.029, and
0.037, from left to right. The prediction in the top row (black line)
matches the data well, even in the rightmost column.}
\label{rho_J}
\end{figure}

Equation (\ref{hJ}b) does more than just predict a lack of
$N$-dependence; it also provides a functional relationship between the
pairwise couplings and the normalized pairwise correlations function,
$\rho_{ij}$.
In Figs.~\ref{rho_J}A-C we plot the pairwise couplings, $J_{ij}$,
versus the normalized pairwise correlation coefficient,
$\rho_{ij}$ (blue dots),
along with the prediction from Eq.\ (\ref{hJ}b) (black line).
Consistent with with our
predictions, the data in Figs.~\ref{rho_J}A-C essentially
follows a line -- the one predicted by Eq.~(\ref{hJ}b).

A relationship between the pairwise couplings and the correlations
coefficients has been sought previously, but
for the more standard Pearson correlation 
coefficient \cite{Schneidman05,Yu08,Tang08}.
Our analysis explains why it was not found. The Pearson correlation
coefficient, denoted $c_{ij}$, is given by

\begin{equation}
c_{ij}\equiv \frac{\langle r_i r_j \rangle-\langle r_i \rangle \langle r_j \rangle }
{\big[(\langle r_i^2 \rangle-\langle r_i \rangle^2)(\langle r_j^2
\rangle-\langle r_j \rangle^2)\big]^{1/2}}
\, .
\label{c-ij}
\end{equation}

\noindent
In the small $\langle r_i \rangle$ limit -- the limit of interest --
the right hand side of Eq.~\eqref{c-ij} is approximately equal to
$[\langle r_i \rangle\langle r_j \rangle]^{1/2} \, \rho_{ij}$. Because
$[\langle r_i \rangle\langle r_j \rangle]^{1/2}$ depends on the local
fields, $h_i$ and $h_j$ (see Eq.\ (\ref{hJ}a)) {\em and} there is a
one-to-one relationship between $\rho_{ij}$ and $J_{ij}$
(Eq.~(\ref{hJ}b)), there can't be a one-to-one relationship between
$c_{ij}$ and $J_{ij}$. We verify the lack of a relationship in
Figs.~\ref{rho_J}D-E, where we again plot $J_{ij}$, but this time
versus the standard correlation coefficient, $c_{ij}$. As predicted,
the data in Figs.~\ref{rho_J}D-E is scattered over two
dimensions. This suggests that $\rho_{ij}$, not $c_{ij}$, is the
natural measure of the correlation between two neurons when the have a
binary representation, something that has also been suggested by Amari
based on information-geometric arguments \cite{Amari08}.

Note that the lack of a simple relationship between the pairwise
couplings and the standard correlation coefficient has been a major
motivation in building maximum entropy models
\cite{Schneidman05,Yu08}. This is for good reason: if there is a
simple relationship, knowing the $J_{ij}$'s adds essentially
nothing. Thus, plotting $J_{ij}$ versus $\rho_{ij}$ (but not $c_{ij}$)
is an important test of one's data, and if the two quantities fall on
the curve predicted by Eq.~(\ref{hJ}b), the maximum entropy model is
adding very little information, if any.

\section{Is there anything wrong with using small time bins?}
\label{small-time-bin-wrong}

An outcome of our perturbative approach is that the goodness of fit
measure, $\Delta_N$, decreases linearly with bin size (see
Eq.~\eqref{DeltaN-def}). This suggests that one could make the
pairwise model look better and better simply by making the bin size
smaller and smaller. Is there anything wrong with this? The answer is
yes, for reasons we discussed in Sec.~\ref{problem:extrapolation};
here we emphasize and expand on this issue, as it is an important one
for making sense of experimental results.

The problem arises because what we have been calling the ``true''
distribution is not really the true distribution of spike trains. It
is the distribution assuming independent time bins, an assumption that
becomes worse and worse as we make the bins smaller and smaller.
(We use this potentially confusing nomenclature primarily
because all studies of
neuronal data carried out so far have assumed temporal independence,
and compared the pairwise distribution to the temporally independent
-- but still correlated across neurons -- distribution
\cite{Schneidman05,Shlens06,Yu08,Tang08}. In addition, the correct
name ``true under the assumption of temporal
independence,'' is unwieldy.)
Here we quantify
how much worse. In particular, we show that if one uses time bins that are
small compared to the characteristic correlation time in the spike
trains, the pairwise model will not provide a good description of the
data. Essentially, we show that, when the time bins are too small, the
error one makes in ignoring temporal correlations is larger than the
error one makes in ignoring correlations across neurons.

As usual, we divide time into bins of size $\delta t$. However,
because we are dropping the independence assumption, we 
use $\br^t$, rather than $\br$, to denote the response in bin $t$.
The full probability distribution over all time bins is denoted
$\calP(\br^1, \dots,\br^M)$. Here $M$ is the number of bins; it is
equal to $T/\delta t$ for spike trains of length $T$.

If time bins are approximately independent, we can write
\begin{equation}
\calP(\br^1, \dots,\br^M)\approx \prod_t p_{true}(\br^t)
\, ,
\label{temp-ind}
\end{equation}

\noindent
and if the pairwise model is a good one, we have
\begin{equation}
p_{true}(\br^t)\approx  p_{pair}(\br^t).
\label{pair-approx}
\end{equation}

\noindent
Combining Eqs.\ (\ref{temp-ind}) 
and Eq.\ (\ref{pair-approx}) then gives us an especially simple
expression for the full probability distribution:
{$\calP(\br^1, \dots,\br^M)\approx \prod_t p_{pair}(\br^t)$.

The problem with small time bins lies in Eq.~(\ref{temp-ind}):
the right hand side is a good approximation to the true
distribution} when the time bins are large compared to the spike train
correlation time, but it is a bad approximation when
the time bins are small (because adjacent time bins become
highly correlated). To quantify how bad, we compare the error one makes
assuming independence across time to the error one makes assuming
independence across neurons. The ratio of those two errors, denoted
$\gamma$, is given by
\begin{equation}
\gamma=\frac{D_{KL}\left(\calP(\br^1, \dots,\br^M) \big| \big|
\prod_t p_{pair}(\br^t) \right)}{M D_{KL}(p(\br)||p_{ind}(\br))}
\, .
\label{gamma-def}
\end{equation}

\noindent
It is relatively easy to compute
$\gamma$ in the limit of small time bins (see Sec. \ref{ind-assumption}),
and we find that
\begin{equation}
\gamma =  \Delta_N + (M-1) +
{\log_2 M \over \gpair \delta}
\, . \label{gamma}
\end{equation}

As expected, this reduces to our old result, $\Delta_N$, when there is
only one time bin ($M=1$). When $M$ is larger than 1, however (which
is, of course, the case of interest), the second term is always at
least one, and for small bin size, the third term is much larger than
one. Consequently, if we use bins that are small compared to the
temporal correlation time of the spike train, the pairwise model will
do a very bad job describing the full, temporally correlated
spike trains.

\section{Discussion}

Probability distributions over the configurations of biological
systems are extremely important quantities. However, because of the
large number of interacting elements comprising such systems, these
distributions can almost never be determined directly from
experimental data. Using parametric models to approximate the true
distribution is the only existing alternative. While such models are
promising, they are typically applied only to small subsystems, not
the full system. This raises the question: are they good models of the
full system?.

We answered this question for a class of parametric models known as
pairwise models. We focused on a particular application, neuronal
spike trains, and our main result is as follows: if one were to record
spikes from multiple neurons, use sufficiently small time bins and a
sufficiently small number of
cells, and assume temporal independence, then a
pairwise model would almost always succeed in
matching the true (but temporally independent)
distribution -- whether or not it would match the
true (but still temporally independent)
distribution for large time bins or a large number of cells.
In other words, pairwise
models in the ``sufficiently small'' regime, what we refer to as the
perturbative regime, have almost no predictive value for what will
happen with large populations. This makes extrapolation from small to
large systems dangerous.

This observation is important because pairwise models, and in
particular maximum entropy pairwise models, have recently attracted a
great deal of attention: they have been applied to salamander and
guinea pig retinas \cite{Schneidman05}, primate retina
\cite{Shlens06}, primate cortex \cite{Tang08}, cultured cortical
networks \cite{Schneidman05}, and cat visual cortex \cite{Yu08}.
These studies have mainly operated close
to the perturbative regime.
For example, Schneidman et al.\ \cite{Schneidman05} had $N \nubar
\delta t \approx 0.35$, Tang et al.\ \cite{Tang08} had $N
\nubar \delta t \approx 0.06$ to $0.4$
(depending on the preparation), and Yu et al.\ had $N \nubar
\delta t \approx 0.2$.
For these studies, then, it would be
hard to justify extrapolating to large populations.

The study by Shlens et al.\ \cite{Shlens06}, on the other hand, might
be more amenable to extrapolation. This is because spatially localized
stimuli were used to stimulate retinal ganglion cells, for which a
nearest neighbor maximum entropy models provided a good fit to their
data. (Nearest neighbor means $J_{ij}$ is zero unless cell $i$ and
cell $j$ are adjacent.) As is not hard to show, for nearest neighbor
models the small parameter in the perturbative expansion is $K \nubar
\delta t$ where $K$ is the number of nearest neighbors. Since $K$ is
fixed, independent of the population size, the small parameter will
not change as the population size increases. Thus, Shlens et al.\ may
have tapped into the large population behavior even though they
considered only a few cells at a time in their analysis.

\subsection{Time bins and population size}

That the pairwise model is always good if $N \nubar \delta t$ is
sufficiently small has strong implications: if we want to build a good
model for a particular $N$, we can simply choose a bin size that is
small compared to $1/N \nubar$. However, one of the assumptions in all
pairwise models used on neural data is that bins at different times
are independent. This produces a tension between small time bins and
temporal independence: small time bins essentially ensure that a
pairwise model will provide a close approximation to
a model with independent bins, but
they make adjacent bins highly correlated. Large time bins
come with no such assurance, but they make adjacent bins
independent. Unfortunately, this tension is often unresolvable in
large populations, in the sense that pairwise models are assured
to work only up to populations of size $1/(\nubar \tau_{\rm corr})$
where $\tau_{\rm corr}$ is the typical correlation time. Given that
$\nubar$ is at least several Hz, for experimental paradigms in which
the correlation time is more than a few hundred ms, $1/(\nubar
\tau_{\rm corr})$ is about one, and this assurance does not apply to
even moderately sized populations of neurons. 

These observations are especially relevant for studies that use small
time bins to model spike trains driven by natural stimuli. This is
because the long correlation times inherent in natural stimuli are
passed on to the spike trains, so the assumption of independence
across time (which is required for the independence assumption to be
valid) breaks badly. Knowing that these models are successful in
describing spike trains under the independence assumption, then, does
not tell us whether they
will be successful in describing full, temporally correlated,
spike trains. Note that for studies that
use stimuli with short correlation times (e.g., non-natural stimuli
such as white noise), the temporal correlations in the spike trains
are likely to be short, and using small time bins may be perfectly
valid.

The only study that has investigated the issue of temporal
correlations in maximum entropy models does indeed support the above
picture \cite{Tang08}: for the parameters used in that study
($N \nubar \delta t = 0.06$ to $0.4$), the maximum entropy pairwise
model provided a good fit to the data ($\Delta_N$ was typically
smaller than 0.1), but it did not do a good
job modeling the temporal structure of the spike trains.
 
 \subsection{
Other biological problems that have been approached with pairwise
models, e.g, protein folding
}

As mentioned in the Introduction, in addition to the studies on
neuronal data, studies on protein folding have also emphasized the
role of pairwise interactions \cite{Socolich05,Russ05}. Briefly,
proteins consist of strings of amino acids, and a major question in
structural biology is: what is the probability distribution of amino
acid strings in naturally folding proteins? One way to answer this is
to approximate the full probability distribution of naturally folding
proteins from knowledge of single-site and pairwise distributions. One
can show that there is a perturbative regime for proteins as well.
This can be readily seen using the celebrated HP protein model
\cite{Dill85}), where a protein is composed of only two types of amino
acids: if, at each site, one amino acid type is preferred and occurs
with high probability, say $1-\delta$ ($\delta \ll 1$), then a protein of
length shorter than $1/\delta$ will be in the perturbative regime,
and, therefore, a good match between the true distribution and the
pairwise distribution for such a protein is virtually guaranteed.

Fortunately, though, the properties of real proteins generally prevent
this from happening: at the majority of sites in a protein, the
distribution of amino acids is {\em not} sharply peaked around one
amino acid. Even for those sites that are sharply peaked  (the
evolutionarily-conserved sites), the probability of the most likely
amino acid rarely exceeds $90\%$ \cite{Lockless99,Vargas-Madraz94}.
This puts proteins consisting of only a few amino acids out of the
perturbative regime, and puts longer proteins -- the ones usually
studied using pairwise models -- well out of it.

This difference is fundamental: because many of the studies that have
been carried out on neural data were in the perturbative regime, the
conclusions of those studies -- specifically, the conclusion that
pairwise models provide accurate descriptions of large populations of
neurons -- is not yet supported. This is not the case for the protein
studies, because they are not in the perturbative regime. Thus, the
evidence that pairwise models provide accurate descriptions of protein
folding remain strong and exceedingly promising.
 
 \subsection{Outlook}

We have developed a framework for assessing the validity of pairwise
models applied to small systems. Essentially, we developed a set 
of tests to determine whether one's
data is in the perturbative regime, a regime in which extrapolation to
large populations is not warranted. This should serve as a useful
guide, not just for analyzing experiments, but also for designing
them.

Although our framework is general, we focused primarily on its
application to neural data. One of our main results is that
the bin size carries an important tradeoff: if the bin size is small,
then pairwise models work well, but at the price of ignoring temporal
correlations; if the bin size is large enough so that adjacent bins
are weakly correlated, then there is no guarantee that pairwise models
will work at all. Pairwise models with small time bins,
therefore, might be rescued by a
small modification: take into account correlations across time as well
as neurons. This would increase the complexity of the models,
but the amount of data one needs to fit them would still not be so
large, as
only pairwise correlations and single neuron firing rates need
to be estimated.
Whether this modification would produce good models is
not clear, but if it did it would bring us much closer to a
fundamental understanding of neural systems.

\section{Methods}

\subsection{The behavior of the true entropy in the large $N$ limit}
\label{s_asymptotic}

To understand how the true entropy behaves in the large $N$ limit,
we need only
express the difference of the entropies as a mutual information.
Using $S_N$ to denote the true entropy of $N$ neurons and $I(1;N)$ to
denote the mutual information between one neuron and the other $N$
neurons in a population of size $N+1$, we have
\begin{equation}
(S_N + S_1) - S_{N+1} = I(1;N) \ \ \ \Rightarrow \ \ \ S_{N+1} - S_N
= S_1 - I(1;N)
\, .
\label{deltasdn}
\end{equation}

\noindent
If knowing the activity of $N$ neurons does not constrain the firing of
neuron $N+1$, then the single neuron entropy, $S_1$, will exceed the
mutual information, $I(1;N)$, and the entropy will be an increasing
function of $N$. For the entropy to be linear in $N$, all we need is
that the mutual information saturates with $N$. Because of synaptic
failures, this is a reasonable assumption for
networks of neurons: even if we observed all the neurons, there is
still residual noise associated with uncertainty about which vesicles
release neurotransmitter. Thus, using $I(1;\infty)$ to denote the asymptotic
value of the mutual information and $\langle S_1 \rangle$ to denote
the average single-neuron entropy, we have
\begin{equation}
S_N = N[\langle S_1 \rangle - I(1;\infty)] + {\rm corrections}
\, ,
\label{dsdn}
\end{equation}

\noindent
where the corrections are sublinear in $N$.

\subsection{Perturbative Expansion}
\label{perturbative_expansion}

Our main quantitative result, given in Eq.~\eqref{KLs_N}, is that the
KL divergence between the true distribution and both the independent
and pairwise distributions can be computed perturbatively as an
expansion in powers of $N \delta$. Here we carry out this
expansion, and derive explicit expressions for the quantities $g_{ind}$
and $g_{pair}$.

To simplify our notation, here we use $p(\br)$ for the true
distribution. The critical step in computing the KL divergences
perturbatively is to use the Sarmanov-Lancaster
expansion \cite{Sarmonov62,Sarmonov63,Lancaster58,Lancaster58,Lancaster63,Bahadur61} for
$p(\br)$,
\begin{equation}
p(\br) = p_{ind}(\br) \, (1+\xi_p(\br) )
\label{p}
\end{equation}

\noindent
where
\begin{subequations}
\begin{align}
p_{ind}(\br) & = \frac{\exp \sum_i {\calH}^p_i r_i }
{\prod_i \left[ 1 + \exp (\calH^p_i r_i) \right]}
\\
\xi_p(\br) &
\equiv \sum_{i<j} \calJ^p_{ij} \delta r_i \delta r_j + \sum_{i<j<k}
\calK^p_{ijk}  \delta r_i \delta r_j \delta r_k + \dots
\\
\delta r_i & \equiv  r_i- \barr_i
\\
\barr_i & \equiv  (1+\exp(-\calH_i))^{-1}
\, .
\end{align}
\label{p_expansion}
\end{subequations}

This expansion has a number of important, but not immediately obvious,
properties. First, as can be shown by a direct calculation,
\begin{equation}
\langle r_i \rangle_p =
\langle r_i \rangle_{ind} =
\barr_i
\label{ribar}
\end{equation}

\noindent
where the subscripts $p$ and $ind$ indicate an average with respect to
$p(\br)$ and $p_{ind}(\br)$, respectively. This has an immediate
corollary,
\begin{equation}
\langle \delta r_i \rangle_{ind} = 0
\, .
\end{equation}

\noindent
This last relationship is important, because it tells us that
if a product of $\delta r$'s contains any terms linear in one of the
$\delta r_i$, the whole product averages to zero under the independent
distribution. This can be used to show that
\begin{equation}
\langle \xi_p(\br) \rangle_{ind} = 0
\label{xibar}
\end{equation}

\noindent
from which it follows that
\begin{equation}
\sum_\br p(\br) = \langle (1 + \xi_p(\br) \rangle_{ind} = 1
\, ,
\end{equation}

\noindent
which tells us that $p(\br)$ is properly normalized.
\noindent
Finally, a slightly more involved calculations provides us with a
relationship between the parameters of the model and the moments:
for $i \ne j \ne k$,

\begin{subequations}
\begin{align}
\langle \delta r_i \delta r_j \rangle_p & =
\barr_i (1-\barr_i )\barr_j (1-\barr_j) \calJ^p_{ij}
\\
\langle \delta r_i \delta r_j \delta r_k \rangle_p &=
\barr_i (1-\barr_i )\barr_j (1-\barr_j)\barr_k (1-\barr_k)
\calK^p_{ijk}
\, .
\end{align}
\label{moments}
\end{subequations}

\noindent
Virtually identical expressions hold for higher order moments.
It is this last set of relationships that make the
Sarmanov-Lancaster expansion
so useful.

Note that Eqs.~(\ref{moments}a) and (\ref{moments}b), along with the
expression for the normalized correlation coefficients given in
Eq.~\eqref{rho_norm}, imply that

\begin{subequations}
\begin{align}
(1-\barr_i ) (1-\barr_j) \calJ^p_{ij} &= \rho^p_{ij}
\\
(1-\barr_i ) (1-\barr_j) (1-\barr_k) \calK^p_{ijk} &= \rho^p_{ijk}
\, .
\end{align}
\label{jk_moments}
\end{subequations}

\noindent
These identities will be extremely useful for simplifying expressions
later on.

Because the moments are so closely related to the parameters of
the distribution, moment matching is especially convenient: to
construct a distribution whose moments match those of $p(\br)$ up to some
order, one simply needs to ensure that the parameters of that
distribution, $\calH_i$, $\calJ_{ij}$, $\calK_{ijk}$, etc., are identical to
those of the true distributions up to the order of interest.
In particular, let us write down a new distribution, $q(\br)$,

\begin{subequations}
\begin{align}
q(\br) & = p_{ind}(\br)
\, (1+\xi_q(\br) )
\\
\xi_q(\br) &
=\sum_{i<j} \calJ^q_{ij} \delta r_i \delta r_j + \sum_{i<j<k}
\calK^q_{ijk}  \delta r_i \delta r_j \delta r_k + \dots
\, .
\end{align}
\label{q_expansion}
\end{subequations}

\noindent
We can recover the independent distribution by
letting $\xi_q(\br)=0$, and we can recover the pairwise distribution by
letting $\calJ^q_{ij} = \calJ^p_{ij}$. This allows us
to compute $D_{KL}(p||q)$ in the general case, and then either set
$\xi_q$ to zero or set $\calJ^q_{ij}$ to $\calJ^p_{ij}$.

Note that expressions analogous to those in
Eq.~(\ref{xibar}-\ref{jk_moments})
exist for averages with respect to $q(\br)$; the only difference is
that $p$ is replaced by $q$.

\subsubsection{The KL divergence in the Sarmanov-Lancaster representation}
\label{section:KL}

Using Eqs.~\eqref{p} and (\ref{q_expansion}a) and a small amount of
algebra, the KL
divergence between $p(\br)$ and $q(\br)$ may be written
\begin{equation}
D_{KL}(p||q)=
\frac{1}{\ln 2} \left\langle f(\xi_p(\br), \xi_q(\br))\right\rangle_{ind}
\label{fxybar}
\end{equation}

\noindent
where
\begin{equation}
f(x, y) \equiv (1+x)[ \ln(1+x) - \ln(1+y)] - (x-y)
\, .
\label{fxy}
\end{equation}

\noindent
To derive Eq.~\eqref{fxybar}, we used the fact that
$\langle \xi_p \rangle_{ind}
=\langle \xi_q \rangle_{ind} = 0$ (see Eq.~\eqref{xibar}).
The extra term $(x-y)$ was included to ensure that
$f(x,y)$ and its first derivatives vanish at $x=y=0$, something that
greatly simplifies our analysis later on.

Our approach is to Taylor expand the right hand side of
Eq.~\eqref{fxybar} around $\xi_p =
\xi_q = 0$, compute each term, and then sum the {\em whole} series. Using
$a_{nm}$ to denote the coefficients of the Taylor series, we have
\begin{equation}
D_{KL}(p||q)= \frac{1}{\ln 2}
\sum_{mn} a_{mn} \left\langle \xi_p(\br)^m \xi_q(\br)^n
\right\rangle_{ind}
\, .
\label{taylor}
\end{equation}

\noindent
Note that because $f(x,y)$ and its first derivatives vanish
at $x=y=0$, all terms in this sum have $m+n \ge 2$.

Because both $\xi_p$ and $\xi_q$ are themselves sums, an exact
calculation of the terms in Eq.~\eqref{taylor} would be difficult.
However, as we show in Sec.~\ref{section:averages} (in particular
Eqs.~\eqref{xipq2_sum} and \eqref{xipq3_sum}), they can be
computed to lowest order in $N\delta$, and the result is
\begin{eqnarray}
\langle \xi_p(\br)^m \xi_q(\br)^n \rangle_{ind} & = &
\frac{1}{\ln 2}
\sum_{i < j}
\barr_i \barr_j (\rho^p_{ij})^m (\rho^q_{ij})^n
+
\left[ \barr_j (-\barr_i \rho^p_{ij})^m (-\barr_i \rho^q_{ij})^n
+
\barr_i \leftrightarrow \barr_j
\right]
\ \ \ 
\nonumber
\\
& + &
\frac{1}{\ln 2}
\sum_{i < j < k}
\barr_i \barr_j \barr_k
(\rhothree^p_{ijk})^m
(\rhothree^q_{ijk})^n
+ \order \left((N\delta)^4\right)
\label{xipq_final}
\end{eqnarray}

\noindent
where $\rhothree^p_{ijk}$ and $\rhothree^q_{ijk}$ are given by
\begin{equation}
\rhothree^x_{ijk} \equiv \rho^x_{ijk}
+ \rho^x_{ij} + \rho^x_{ik} + \rho^x_{jk}
=
\frac{\langle r_i r_j r_k \rangle_x - \barr_i \barr_j \barr_k}
{\barr_i \barr_j \barr_k}
\, ,
\label{rhothree} 
\end{equation}

\noindent
$x=p,q$. The last equality in Eq.~\eqref{rhothree}
follows from a small amount of
algebra and the definition of the correlation coefficients given in
Eq.~\eqref{rho_norm}. Equation \eqref{xipq_final} is valid only
when $m+n \ge 2$, which is the case of interest to us (since the
Taylor expansion of $f(x,y)$ has only terms with $m+n \ge 2$).

The important point about Eq.~\eqref{xipq_final} is that the $m$ and $n$
dependence follows that of the original Taylor expansion. Thus, when
we insert this equation back into Eq.~\eqref{taylor}, we recover our original
function -- all we have to do is interchange the sums. For example,
consider inserting the first term in Eq.~\eqref{xipq_final} into
Eq.~\eqref{taylor},
\begin{equation}
\sum_{m,n} a_{mn} \sum_{i<j} \barr_i \barr_j (\rho^p_{ij})^m (\rho^q_{ij})^n
= \sum_{i<j} \barr_i \barr_j \sum_{m,n} a_{mn} (\rho^p_{ij})^m (\rho^q_{ij})^n
= \sum_{i<j} \barr_i \barr_j f(\rho^p_{ij}, \rho^q_{ij})
\, .
\nonumber
\end{equation}

\noindent
Performing the same set of manipulations on all of
Eq.~\eqref{xipq_final} leads
to

\begin{eqnarray}
\langle \xi_p(\br)^m \xi_q(\br)^n \rangle_{ind} & = &
\frac{1}{\ln 2}
\sum_{i < j}
\barr_i \barr_j f(\rho^p_{ij}, \rho^q_{ij})
+
\barr_j f(-\barr_i \rho^p_{ij}, -\barr_i \rho^q_{ij})
+
\barr_i f(-\barr_j \rho^p_{ij}, -\barr_j \rho^q_{ij})
\ \ \ 
\nonumber
\\
& + &
\frac{1}{\ln 2}
\sum_{i < j < k}
\barr_i \barr_j \barr_k
f(\rhothree^p_{ijk}, \rhothree^q_{ijk})
+ \order \left(N\delta)^4\right)
\, .
\label{xipq}
\end{eqnarray}

This expression is true in general (except for some technical
considerations; see Sec.~\ref{section:averages}); to restrict it to the KL
divergences of interest we set $p(\br)$ to $p_{true}(\br)$ and
$q(\br)$ to either $p_{ind}(\br)$ or $p_{pair}(\br)$. In the first
case, $q(\br) = p_{ind}(\br)$, $\xi_q(\br) = 0$, which in term implies
that $\calJ^q_{ij}=0$, and thus $\rho^q_{ij} = 0$.
Using Eq.~\eqref{xipq}, we have (to lowest nonvanishing order in
$N\delta$),

\begin{equation}
D_{KL}(p_{true}||p_{ind})=
\frac{1}{\ln 2}
\sum_{i < j} \barr_i \barr_j f(\rho^p_{ij}, 0)
+ \order\left((N \delta)^3\right)
\, .
\label{dkl_ind}
\end{equation}

\noindent
Then, defining
\begin{equation}
\gind \equiv \frac{1}{N(N-1)\ln(2)}\sum_{i<j}
\frac{\barr_i}{\delta}\frac{\barr_j}{\delta} f(\rho^p_{ij}, 0)
\label{gind}
\, ,
\end{equation}

\noindent
and recalling that $\delta = \nubar \delta t$, we see that
Eq.~\eqref{dkl_ind} is equivalent to Eq.~\eqref{KLs_Na}.

In the second case, $q(\br) = p_{pair}(\br)$, the first and second
moments of $p_{pair}(\br)$ and $p_{true}(\br)$ are equal. This in turn
implies, using Eq.~\eqref{moments}, that $\calJ^q_{ij}=\calJ^p_{ij}$,
and thus $\rho^p_{ij}=\rho^q_{ij}$. Because $f(x,x)=0$ (see
Eq.~\eqref{fxy}), we see that the first three terms on the right hand
side of Eq.~\eqref{xipq} -- those involving $i$ and $j$ but not
$k$ -- vanish. The next order term does not vanish, and yields

\begin{equation}
D_{KL}(p_{true}||p_{pair}) =
\frac{1}{\ln 2}
\sum_{i < j < k} \barr_i \barr_j \barr_k
f(\rhothree^p_{ijk}, \rhothree^q_{ijk})
+ \order \left( (N\delta)^4 \right)
\, .
\label{dkl2_contracted}
\end{equation}

\noindent
Defining
\begin{equation}
\gpair \equiv
\frac{1}{N(N-1)(N-2)\ln(2)}\sum_{i<j<k} \frac{r_i}{\delta}
\frac{r_j}{\delta} \frac{r_k}{\delta}
f(\rhothree^p_{ijk},\rhothree^q_{ijk})
\, ,
\label{gpair}
\end{equation}
\noindent
we see that Eq.~\eqref{dkl2_contracted} reduces to Eq.~\eqref{KLs_Nb}.

\subsubsection{Local fields, pairwise couplings and moments}
\label{local-fields}

In this section we derive, to leading order in $N\delta$, expressions
relating the local fields and pairwise couplings of the maximum
entropy model, $h_i$ and $J_{ij}$, to the first and second moments.
These are the expressions reported in Eq.~\eqref{hJ}.

To do this, we simply compute the first and second moment under the
assumption that $N\delta$ is small. This calculation proceeds along
the same lines as in the previous section, with one extra
consideration: the
quadratic term in the maximum entropy distribution,
Eq.~\eqref{maxent}, is proportional to $r_i r_j$, not $\delta r_i
r_j$. However, to lowest order in $N\delta$, this doesn't matter.
That's because
\begin{equation}
\sum_{i < j} J_{ij} r_i r_j = \sum_{i < j} J_{ij} \delta r_i \delta r_j
+ r_i \sum_{j \ne i} J_{ij} \barr_j
+ \hbox{constants}
\, .
\nonumber
\end{equation}

\noindent
where $\barr_i$ is defined as in Eq.~(\ref{p_expansion}d) except
with $\calH^p_i$ replaced by $h_i$, and we used the fact that
$J_{ij} = J_{ji}$. The second term
introduces a correction to the local fields, $h_i$. However, the
correction is $\order(N\delta)$, so we drop it. We should keep in
mind, though, that our final expression for $h_i$ will have
corrections of $\order(N\delta)$.

Using Eq.~\eqref{maxent}, but with $r_i$ replaced by $\delta r_i$
where it appears with $J_{ij}$, we may write (after a small amount of
algebra)
\begin{equation}
p_{maxent}(\br) = p_{ind}(\br)
\frac{1 + \xi_x(\br) + \psi(\xi_x(\br))}
{1 + \langle \xi_x(\br) + \psi(\xi_x(\br)) \rangle_{ind}}
\label{p-me}
\end{equation}

\noindent
where $p_{ind}(\br)$ is the same as the function $p_{ind}(\br)$
defined in Eq.~(\ref{p_expansion}a), except that $\calH^p_{ij}$ is
replaced by $h$, the subscript ``{\em ind}'' indicates, as usual, an
average with respect to $p_{ind}(\br)$, and the two functions
$\xi_x(\br)$ and $\psi(x)$ are given by
\begin{equation}
\xi_x(\br) \equiv
\sum_{i < j} J_{ij} \delta r_i \delta r_j
\label{xi_x}
\end{equation}

\noindent
and
\begin{equation}
\psi(x) \equiv e^x - 1 - x
\, .
\label{psi}
\end{equation}

Given this setup, we can use
Eqs.~\eqref{phi_g1} and \eqref{phi_g2} to compute the moments under
the maximum entropy model. That's because
both $\psi(x)$ and its first derivative vanish at $x=0$, which
are the two conditions required for
these equations to be valid. Using also the
fact that $\langle \delta r_i \rangle_{ind} = 0$,
Eqs.~\eqref{phi_g1} and \eqref{phi_g2} imply that
\begin{subequations}
\begin{align}
\langle \xi_x(\br) + \psi(\xi_x(\br)) \rangle_{ind} & =
\sum_{i < j} \barr_i \barr_j \psi(J_{ij})
+ \order \left( (N\delta)^3 \right)
\\
\langle r_i \rangle_{maxent} & = \left(1 + \exp(-h_i) \right)^{-1}
+ \order \left( N\delta^2 \right)
\\
\langle \delta r_i \delta r_j \rangle_{maxent} &=
\barr_i \barr_j [ \psi(J_{ij}) + J_{ij}]
+ \order \left( N\delta^3 \right)
\, .
\end{align}
\label{moments-me}
\end{subequations}

\noindent
The term ``$+J_{ij}$'' in Eq.~(\ref{moments-me}c) came from
$\langle \delta r_i \delta r_j \xi_x(\br) \rangle_{ind}$;
see numerator in Eq.~\eqref{p-me}.
Note that for the second two equations, we used the fact that, to
lowest order in $N\delta$, the
denominator in Eq.~\eqref{p-me} is equal to 1.

Finally, using Eq.~\eqref{norm-corr} for the normalized correlation
coefficient, dropping the subscript ``{\em maxent}'' (since the
first and second moments are the same under the maxent and true
distributions), and inverting Eqs.~(\ref{moments-me}b) and
(\ref{moments-me}c) to express the local fields and
coupling coefficients in terms of the first and second moments, we
arrive at Eq~\eqref{hJ}.

\subsubsection{Averages of powers of $\xi_p$ and $\xi_q$}
\label{section:averages}

Here we compute $\langle \xi_p^m \xi_q^n \rangle_{ind}$, which, as can
be seen in Eq.~\eqref{taylor}, is the the key quantity in our
perturbation expansion. Our starting point is to (formally) expand the
sums that make up $\xi_p$ and $\xi_q$ (see
Eqs.~(\ref{p_expansion}b) and ~(\ref{q_expansion}b)), which yields

\begin{equation}
\langle \xi_p(\br)^m \xi_q(\br)^n \rangle_{ind} =
\sum_{l=2}^\infty
\sum_{\{m_1, \dots, m_l\}}
\psi_{m_1, \dots, m_l}^{(l)}
\sum_{i_1 < \dots < i_l}
\langle \delta r_{i_1}^{m_1} \dots, \delta r_{i_l}^{m_l}
\rangle_{ind}
\, .
\label{xipq_sum}
\end{equation}

\noindent
The sum over ${\{m_1, \dots, m_l\}}$ is a sum
over all possible configurations of the $m_i$. The
coefficient $\psi^{(l)}_{m_1, \dots, m_l}$ are complicated functions
of the $\calJ^p_{ij}, \calJ^q_{ij}, \calK^p{ijk}, \calK^q_{ijk}$, etc.
Computing these functions is straightforward, although somewhat
tedious, especially when $l$ is large;
below we compute them only for $l=2$ and 3. The reason $l$
starts at 2 is that $m+n \ge 2$; see Eq.~\eqref{taylor}.

What we will show is that all terms with superscript $(l)$ are
$\order(\delta^l)$. To do this, we first note that, because
the right hand
side of Eq.~\eqref{xipq_sum} is an average with respect to the independent
distribution, the average of the product is the product of the
averages,

\begin{equation}
\langle \delta r_{i_1}^{m_1} \delta r_{i_2}^{m_2} \dots, \delta r_{i_l}^{m_l}
\rangle_{ind}
=
\langle \delta r_{i_1}^{m_1} \rangle_{ind} \langle \delta
r_{i_2}^{m_2} \rangle_{ind} \dots, \langle \delta r_{i_l}^{m_l} \rangle_{ind}
\, .
\label{drmm}
\end{equation}

\noindent
Then, using the fact that $\delta r_i=(1-\barr_i)$ with probability $\barr_i$
and $\delta r_i=(1-\barr_i)$ with probability $(1-\barr_i)$  (see Eq.~(\ref{p_expansion}c)), we have 
%
%
\begin{equation}
\langle \delta r_i^m \rangle_{ind} = \barr_i
(1-\barr_i)^m+(1-\barr_i)(-\barr_i)^m
= \barr_i (1 - \barr_i)^m \left[ 1 -
\left(\frac{-\barr_i}{1-\barr_i} \right)^{m-1}
\right]
\label{drm}
\, .
\end{equation}

\noindent
The significance of this expression is that, for $m > 1$,
$\langle \delta r_i^m \rangle_{ind} \sim \order(r_i) \sim \order(\delta)$,
independent
of $m$. Consequently, if all the $m_i$ in Eq.~\eqref{drmm} are greater
than 1, then the right hand side is $\order(\delta^l)$. This shows
that, as promised above,
the superscript $(l)$ labels the
order of the terms.

As we saw in Sec.~\ref{section:KL}, we need to go to third order in
$\delta$, which means we need to compute the terms on the right hand
side of Eq.~\eqref{xipq_sum} with $l=2$ and 3. Let us start with $l=2$,
which picks out only on those terms with two unique
indices. Examining the expressions for $\xi_p$ and $\xi_q$ given in
Eqs.~(\ref{p_expansion}b) and (\ref{q_expansion}b), we see
that we must keep only terms involving
$\calJ_{ij}$, since $\calK_{ijk}$ has
three indices, and higher order terms have more. Thus, the $l=2$
contribution to the average in Eq.~\eqref{xipq_sum}, which we denote
$\langle \xi_p(\br) \xi_q(\br) \rangle_{ind}^{(2)}$, is given by

\begin{equation}
\langle \xi_p(\br)^m \xi_q(\br)^n \rangle^{(2)}_{ind} =
\sum_{i < j}
\left\langle
\left(\calJ^p_{ij} \delta r_i \delta r_j \right)^m
\left(\calJ^q_{ij} \delta r_i \delta r_j \right)^n
\right\rangle_{ind}
\, .
\nonumber
\end{equation}

\noindent
Pulling $\calJ^p_{ij}$ and $\calJ^q_{ij}$ out of the averages,
using Eq.~(\ref{jk_moments}a)
to eliminate $\calJ^p_{ij}$ and $\calJ^q_{ij}$ in favor of
$\rho^p_{ij}$ and $\rho^q_{ij}$, and applying
Eq.~\eqref{drm} (while throwing away some of the terms in that
equation that are
higher than second order in $\delta$), the above expression may be
written

\begin{equation}
\langle \xi_p(\br)^m \xi_q(\br)^n \rangle^{(2)}_{ind} =
\sum_{i < j}
\barr_i \barr_j
(\rho^p_{ij})^m
(\rho^q_{ij})^n
\left[1-(-\barr_i)^{m+n-1}-(-\barr_i)^{m+n-1}\right]
\, .
\label{xipq2_sum}
\end{equation}

\noindent
Note that we were not quite consistent in our
ordering with respect to $\delta$, in the sense that we kept some
higher order terms and not others. We did this so that we could
use $\rho_{ij}$ rather than $\calJ_{ij}$, as the former
are directly observable.

For $l=3$ the calculation is more involved, but not substantially so.
Including terms with exactly three unique indices in the sum on the
right hand side of Eq.~\eqref{xipq_sum} gives us

\begin{eqnarray}
\langle \xi_p(\br)^m \xi_q(\br)^n \rangle^{(3)}_{ind} & = &
\sum_{i < j < k}
\left\langle
\left(\calK^p_{ijk} \delta r_i \delta r_j \delta r_k
+ \calJ^p_{ij} \delta r_i \delta r_j
+ \calJ^p_{ik} \delta r_i \delta r_k
+ \calJ^p_{jk} \delta r_j \delta r_k
\right)^m
\right.
\label{xipq3_sum}
\\
\nonumber
& &
\ \ \ \ \ \ \ \ \
\left.
\left(\calK^q_{ijk} \delta r_i \delta r_j \delta r_k
+ \calJ^q_{ij} \delta r_i \delta r_j
+ \calJ^q_{ik} \delta r_i \delta r_k
+ \calJ^q_{jk} \delta r_j \delta r_k
\right)^n
\right\rangle_{ind}
\, .
\ \ \ \ \ \
\end{eqnarray}

This expression is not quite correct, since some of its terms contain
only two unique indices -- these are the terms proportional to
$(\calJ^p_{ij})^m(\calJ^p_{ij})^n$ -- whereas it should contain only
terms with exactly three unique indices. Fortunately, this turns out
not to matter, for reasons we discuss at the end of the section.

To perform the averages in Eq.~\eqref{xipq3_sum}, we would need to
use multinomial expansions, and then
average over the resulting powers of $\delta r$'s.
For the latter, we can work to
lowest order in the $\delta r_i$, which means we only take the first
term in Eq.~\eqref{drm}. This
amounts to replacing every $\delta r_i$ with $1-\barr_i$ (and
similarly for $j$ and $k$),
and in addition multiplying the whole expression by an overall
factor of $\barr_i \barr_j \barr_k$. For example, if $m=1$ and $n=2$,
one of the terms in the multinomial expansion is
$\calK^p_{ijk} \calJ^q_{ij} \calJ^q_{ik}
\langle \delta r_i^3 \delta r_j^2 \delta r_k^2 \rangle_{ind}$.
This average would yield, using Eq.~\eqref{drm} and considering only
the lowest order term,
$\barr_i \barr_j \barr_k (1-\barr_i)^3 (1-\barr_j)^2 (1-\barr_k)^2$.

This procedure also is not quite correct, since
terms with only one factor of $\delta r_i$, which average to zero,
are replaced with $1-\barr_i$. This also turns out not to matter;
again, we discuss why at the end of the section.

We can, then, go ahead and use the above ``replace blindly''
algorithm. Note that the factors of $1-\barr_i$, $1-\barr_j$ and
$1-\barr_k$ turn $\calJ_{ij}$ and $\calK_{ijk}$
into normalized correlation coefficients (see
Eq.~\eqref{jk_moments}), which considerably simplifies our equations.
Using also Eq.~\eqref{rhothree} for $\tilde{\rho}_{ijk}$,
Eq.~\eqref{xipq3_sum} becomes

\begin{equation}
\langle \xi_p(\br)^m \xi_q(\br)^n \rangle^{(3)}_{ind} =
\sum_{i < j < k}
\barr_i \barr_j \barr_k
(\rhothree^p_{ijk})^m
(\rhothree^q_{ijk})^n
\, .
\label{xipq3_loc}
\end{equation}

\noindent
We can now combine Eqs.~\eqref{xipq2_sum} and \eqref{xipq3_loc}, and
insert them into Eq.~\eqref{xipq_sum}. This gives us the first
two terms in the perturbative expansion of
$\langle \xi_p(\br)^m \xi_q(\br)^n \rangle_{ind}$; the result is
written down in Eq.~\eqref{xipq_final} above.

Why can we ignore the overcounting associated with terms in which an
index appears exactly zero or one times? We clearly can't do this in
general, because for such terms, replacing $\delta r_i$ with
$1-\barr_i$ fails -- either because the terms didn't exist in the
first place (when one of the indices never appeared) or because they
averaged to zero (when an index appeared exactly once). In our case,
however, such terms do not appear in the Taylor expansion. To
see why, note first of all that, because of the 
form of
$f(x,y)$, its Taylor expansion
can be written $(x-y)^2 \tilde{f}(x,y)$ where
$\tilde{f}(x,y)$ is finite at $x=y$ (see Eq.~\eqref{fxy}).
Consequently, the expression inside the sum over $i, j$ and $k$
in Eq.~\eqref{xipq3_sum} should really contain a multiplicative factor
that arises from $(\xi_p-\xi_q)^2$, and thus has the form

\begin{equation}
\left((\calK^p_{ijk} - \calK^q_{ijk}) \delta r_i \delta r_j \delta r_k
+ (\calJ^p_{ij} - \calJ^q_{ij}) \delta r_i \delta r_j
+ (\calJ^p_{ik} - \calJ^q_{ik}) \delta r_i \delta r_k
+ (\calJ^p_{jk} - \calJ^q_{jk}) \delta r_j \delta r_k
\right)^2
\, .
\nonumber
\end{equation}

As we saw in the previous section, we are interested in the
third order term only to compute $D_{KL}(p_{true}||p_{pair})$, for
which $\calJ^p_{ij} = \calJ^q_{ij}$. Therefore, the above
multiplicative factor reduces to
$(\calK^p_{ijk} - \calK^q_{ijk})^2 (\delta r_i \delta r_j \delta r_k)^2$.
It is that last factor of $(\delta r_i \delta r_j \delta r_k)^2$ that
is important, since it guarantees that every term in the Taylor
expansion will have all indices appearing at least twice.
Therefore, although Eq.~\eqref{xipq3_sum} is not true in general, it
is valid for our analysis.

We end this section by pointing out that there is a very simple
procedure for computing averages to second order in $\delta$. Consider
a function $\phi(\xi_p, \xi_q)$ such that
$\phi(\xi_p, \xi_q)$ and its first derivatives vanish at
$\xi_p=\xi_q = 0$. Then, based on the above analysis, we have
\begin{equation}
\langle \phi(\xi_p, \xi_q) \rangle_{ind}
= \sum_{i < j} \barr_i \barr_j \phi(\calJ^p_{ij}, \calJ^q_{ij})
+ \order \left((N\delta)^3 \right)
\, .
\label{phi_g1}
\end{equation}

Two easy corollaries of this are: for $k$ and $l$ positive integers,
\begin{subequations}
\begin{align}
\langle \delta r_i^k \phi(\xi_p, \xi_q) \rangle_{ind}
& =
\sum_{j \ne i} \barr_i \barr_j \phi(\calJ^p_{ij}, \calJ^q_{ij})
+ \order \left(N^2\delta^3 \right)
\\
\langle \delta r_i^k \delta r_j^l \phi(\xi_p, \xi_q) \rangle_{ind}
& =
\barr_i \barr_j \phi(\calJ^p_{ij}, \calJ^q_{ij})
+ \order \left(N\delta^3 \right)
\end{align}
\label{phi_g2}
\end{subequations}

\noindent
where the sum in Eq.~(\ref{phi_g2}a) run over $j$ only, and we used
the fact that both $\calJ^p_{ij}$ and $\calJ^q_{ij}$ are symmetric
with respect to the interchange of $i$ and $j$.

\subsection{Generating synthetic data}
\label{Generating_syn_data}

As can be seen in Eq.~\eqref{ptrue}, synthetic data depends
on three sets of parameters:
$\htrue_i, \jtrue_{ij}$, and $\ktrue_{ijk}$. Here we describe how
they were generated.

Our first step was to generate the $\htrue_i$. To do that, we chose a
vector $\br^*=[r^*_1 \dots r^*_{N^*}]$ (where, recall, $N^* = 15$ is
the number of neurons in our base distribution), from an exponential
distribution with mean 0.02. From this we chose the local field
according to Eq.~(\ref{hJ}a),
\begin{equation}
\htrue_i=-\log\left(\frac{1}{r^*_i}-1\right).
\nonumber
\end{equation}

\noindent
In the perturbative regime, a distribution generated with these values
of the local fields will have firing rates approximately equal to the
$r^*_i$; see Eq.~(\ref{hJ}a) and Fig.~\ref{h_J_N}.
We then draw $\jtrue_{ij}$ and $\ktrue_{ijk}$ from Gaussian
distributions with means equal to $0.05$ and $0.02$ and standard
deviations of $0.8$ and $0.5$, respectively. Using non-zero values for
$\calK_{ijk}$, means that the distribution is not pairwise.

\subsection{Bin size and the correlation coefficients}
\label{Bin-size} 

One of our main claims is that $\Delta_N$ is linear in bin size,
$\delta t$. This is true, however, only if $\gind$ and $\gpair$ are
independent of $\delta t$, as can be seen from Eq.~\eqref{DeltaN-def}.
In this section we show that independence is satisfied if $\delta t$
is smaller than the typical correlation time of the responses. For
$\delta t$ larger than such correlation times, $\gind$ and $\gpair$ do
depend on $\delta t$, and $\Delta_N$ is no longer linear in $\delta
t$. Note, though, that the correlation time is always
finite, so there will
always be a bin size below which the linear relationship, $\Delta \sim
\delta t$, is guaranteed.

Examining Eqs.~\eqref{gind} and \eqref{gpair}, we see that
$\gind$ and $\gpair$ depend on the normalized correlation
coefficients, $\rho_{ij}$, $\rhothree_{ijk}$ (we drop superscripts,
since our discussion will be generic).
Thus,
to understand how $\gind$ and $\gpair$ depend on bin size, we need to
understand how the normalized correlation coefficients depend on bin
size.

We start with the second order correlation
coefficient, since it is simplest. The corresponding
cross-correlogram, which we denote $C_{ij}(\tau)$, is given by

\begin{equation}
C_{ij}(\tau) = {1 \over \nu_i \nu_j}
\lim_{T \rightarrow \infty} {1 \over T}
\sum_{kl} \delta(t_i^k - t_j^l - \tau)
\label{C}
\end{equation}

\noindent
where $t_i^k$ is the time of the $k^{\rm th}$ spike on neuron $i$ (and
similarly for $t_j^l$) and $\delta(\cdot)$ is the Dirac
$\delta$-function. The normalization in Eq.~\eqref{C} is slightly
non-standard -- more typical is to divide by something with units of
firing rate ($\nu_i$, $\nu_j$ or $(\nu_i \nu_j)^{1/2}$), to give units
of spikes/s. The normalization we use is convenient, however, because
in the limit of large $\tau$, $C_{ij}(\tau)$ approaches one.

It is slightly tedious, but otherwise straightforward, to show that
when $\delta t$ is sufficiently small that only one spike can occur in
a time bin, $\rho_{ij}$ is related to $C_{ij}(\tau)$ via
\begin{equation}
\rho_{ij} = {1 \over \delta t}
\int_{-\delta t}^{\delta t} d \tau \, (1 - |\tau|/\delta t) \,
(C_{ij}(\tau) - 1)
\, .\label{rhoC}
\end{equation}

\noindent
The (unimportant) factor $(1 - |\tau|/\delta t)$ comes from
the fact that the spikes occur at random locations within a bin.

Equation (\ref{rhoC}) has a simple interpretation: $\rho_{ij}$ is the
average height of the central peak of the cross-correlogram relative
to baseline. How strongly $\rho_{ij}$ depends on $\delta t$ is thus
determined by the shape of the cross-correlogram. If it is smooth,
then $\rho_{ij}$ approaches a constant as $\delta t$ becomes small.
If, on the other hand, there is a sharp peak at $\tau=0$, then
$\rho_{ij} \sim 1/\overline{\nu} \delta t = 1/\delta$
for small $\delta t$, so
long as $\delta t$ is larger than the width of the peak. (The factor
of $\overline{\nu}$ included in the scaling is approximate; it is a
placeholder for an effective firing rate that depends on the indices
$i$ and $j$. It is, however, sufficiently accurate for our purposes.)
A similar relationship exists between the third order correlogram and
the correlation coefficient. Thus, $\rhothree_{ijk}$
is also independent of $\delta t$ in the small $\delta
t$ limit, whereas if the central peak is sharp it scales as
$1/\delta^2$. 

The upshot of this analysis is that the shape of the cross-correlogram
has a profound effect on the correlation coefficients and, therefore, on
$\Delta_N$. What is the shape in real
networks? The answer typically depends
on the physical distance between cells. If two neurons
are close, they are likely to receive common input and thus
exhibit a narrow central peak in their
cross-correlogram. If, on the other hand, the neurons are far
apart, they are less likely to receive common input. In this case, the
correlations come from external stimuli, so the central peak tends to
have a characteristic width given by the temporal
correlation time of the stimulus, typically 100s of milliseconds.

Although clearly both kinds of cross-correlograms exist in any single
population of neurons, it is convenient to analyze them separately. We
have already considered networks in which the cross-correlograms were
broad and perfectly flat, so that the correlation coefficients
were strictly independent of bin size. Based on our analysis in
Sec.~\ref{perturbative_expansion}, we
can also consider the opposite
extreme: networks in which the
the cross-correlograms (both second and higher order) among
nearby neurons exhibit sharp peaks
while those among distant neurons are uniformly equal to 1.
In this regime, the correlation coefficients
depend on $\delta t$: as discussed above, the second order ones scale as
$1/\delta$ and the third as
$1/\delta^2$.
This means that the arguments of
$f(\rho_{ij},0)$ and $f(\rhothree^{true}_{ijk}, \rhothree^{pair}_{ijk})$ are
large. From the definition of $f(x,y)$ in Eq.~\eqref{fxy},
in this regime both are approximately linear in
their arguments (ignoring log corrections).
Consequently,
$f(\rho_{ij}, 0) \sim 1/\delta$ and
$f(\rhothree^{true}_{ijk}, \rhothree^{pair}_{ijk}) \sim 1/\delta^2$.
This implies that $\gind$ and $\gpair$ scale as $N \delta$ and $N^2
\delta$, respectively, and so $\Delta_N \sim N$, independent of
$\delta$. Thus, if the bin size is large compared the the correlation
time, $\Delta_N$ will be approximately independent of bin size.

\subsection{Assessing goodness of fit for independence across time assumption}

\label{ind-assumption}

In this section we derive the expression for $\gamma$ given in
Eq.~\eqref{gamma}. Our starting point is its definition,
Eq.~\eqref{gamma-def}. It is convenient to define $\bR$ to be a
concatenation of the responses in $M$ time bins,
\begin{equation}
\bR \equiv (\br^1, \br^2, ..., \br^M)
\end{equation}

\noindent
where, as in Sec.~\ref{small-time-bin-wrong}, the superscript labels
time, so $\calP(\bR)$ is the full, temporally correlated, distribution.

With this definition, we may write the numerator in
Eq.~\eqref{gamma-def} as
\begin{equation}
D_{KL}\Big(\calP(\bR) || \prod_t p_{pair}(\br^t) \Big)
=
- S^M_{true} -
\sum_t \sum_{\br} p^t_{true}(\br) \log_2 p_{pair}(\br)
\label{DKL_P}
\end{equation}

\noindent
where $S_{full}^M$ is the entropy of the full distribution,
$\calP(\bR)$, the last sum follows from a marginalization
over all but one element of $\calP(\bR)$, and $p^t_{true}(\br)$ is
the true distribution at time $\br$. Note that
$p_{pair}(\br)$ is independent of time, since it is computed from a
distribution averaged over all bins. That distribution, which we have
called $p_{true}(\br)$, is given in terms of $p^t_{true}(\br)$ as
\begin{equation}
p_{true}(\br) = \frac{1}{M} \sum_t p^t_{true}(\br)
\, .
\nonumber
\end{equation}

Inserting this definition into Eq.~\eqref{DKL_P}
eliminates the sum over $t$, and replaces it with $M p_{true}(\br)$.
For simplicity we consider the
maximum entropy pairwise model. In this case,
because $p_{pair}(\br)$ is in the exponential family, and the first
and second moments are the same under the true and maximum entropy
distributions, we can replace $p_{true}(\br)$ with $p_{maxent}(\br)$.
Consequently, Eq.~\eqref{DKL_P} becomes
\begin{equation}
D_{KL}\Big(\calP(\bR) || \prod_t p_{pair}(\br^t) \Big) =
M S_{maxent} - S^M_{true}
\nonumber
\, .
\end{equation}

\noindent
This gives us the numerator in the expression for $\gamma$
(Eq.~\eqref{gamma-def}). The denominator,
$D_{KL}(p_{true}||p_{ind})$,
is equal to $S_{ind} - S_{true}$ (see Eq.~\eqref{KL-Ent-ind}). This
leads to
\begin{equation}
\gamma =
{M(S_{maxent} - S_{true})
\over M (S_{ind} - S_{true})}
+
{MS_{true} - S^M_{true}
\over M (S_{ind} - S_{true})}
\, .
\label{gamma-1}
\end{equation}

\noindent
where we added and subtracted $MS_{true}$ to the numerator.

The first term on the right hand side of Eq.~\eqref{gamma-1}
we recognize, from Eqs.~\eqref{DeltaN-def},
\eqref{KL-Ent-ind} and \eqref{KL-Ent-maxent}, as $\Delta_N$.
To cast the second into a reasonable form, we
define $S_{ind}^M$ to be the entropy of the
distribution that retains the temporal correlations within each neuron
but is independent across neurons. Then, adding and subtracting this
quantity to the numerator in Eq.~\eqref{gamma-1}, and also adding and
subtracting $MS_{ind}$, we have
\begin{equation}
\gamma = \Delta_N +
{
(S^M_{ind} - S^M_{true}) -
M(S_{ind}-S_{true}) +
(M S_{ind} - S^M_{ind})
\over
M(S_{ind} - S_{true})}
\, .
\label{gamma-2}
\end{equation}

\noindent
The key observation is that if $M \delta \ll 1$, then
\begin{equation}
S^M_{ind} - S^M_{true} = \gind N(N-1) (M\delta)^2
\, .
\nonumber
\end{equation}

\noindent
Comparing this with Eq.~\eqref{KLs_Na}, we see that
$S^M_{ind} - S^M_{true}$ is a factor of $M^2$ times larger than
$S_{ind} - S_{true}$. We thus have
\begin{equation}
\gamma = \Delta_N + (M-1) +
{
M S_{ind} - S^M_{ind}
\over
M(S_{ind} - S_{true})}
\label{gamma-final}
\, .
\end{equation}

\noindent
Again assuming $M \delta \ll 1$, and defining
$h(x) \equiv -x \log_2 x - (1-x) \log_2(1-x)$, the last term in this
expression may be written
\begin{equation}
M S_{ind} - S^M_{ind}
= M \sum_i h(r_i) - \sum_i h(Mr_i) \approx M \sum_i r_i \log M
= N\delta \, M \log_2 M
\, .
\end{equation}

\noindent
Inserting this into Eq.~\eqref{gamma-final} and using Eq.~\eqref{KLs_Na}
yields Eq.~\eqref{gamma}.

We have assumed here that $M \delta \ll 1$; what happens when $M
\delta \sim 1$, or larger? To answer this, we rewrite
Eq.~\eqref{gamma-1} as
\begin{equation}
\gamma =
\frac{S_{maxent} - S^M_{true}/M}
{\Delta_N}
\, .
\label{gamma-full}
\end{equation}

\noindent
We argue that in general, as $M$ increases, $S^M_{true}/M$
becomes increasingly different from $S_{maxent}$, since the former was
derived under the assumption that the responses at different time bins
were independent. Thus, Eq.~\eqref{gamma} should be considered a lower
bound on $\gamma$.

\bibliography{mybibliography}

\begin{thebibliography}{10}
\providecommand{\url}[1]{\texttt{#1}}
\providecommand{\urlprefix}{URL }
\expandafter\ifx\csname urlstyle\endcsname\relax
  \providecommand{\doi}[1]{doi:\discretionary{}{}{}#1}\else
  \providecommand{\doi}{doi:\discretionary{}{}{}\begingroup
  \urlstyle{rm}\Url}\fi
\providecommand{\bibAnnoteFile}[1]{%
  \IfFileExists{#1}{\begin{quotation}\noindent\textsc{Key:} #1\\
  \textsc{Annotation:}\ \input{#1}\end{quotation}}{}}
\providecommand{\bibAnnote}[2]{%
  \begin{quotation}\noindent\textsc{Key:} #1\\
  \textsc{Annotation:}\ #2\end{quotation}}
\providecommand{\eprint}[2][]{\url{#2}}

\bibitem{Rieke97}
Rieke F, Warland D, de~Ruyter~van Steveninck RR, Bialek W (1997) Spikes:
  exploring the neural code.
\newblock Cambridge, MA: MIT press.
\bibAnnoteFile{Rieke97}

\bibitem{Russ05}
Russ WP, Lowery DM, Mishra P, Yaffe MB, Ranganathan R (2005) Natural like
  function in artificial ww domains.
\newblock Nature 437:579--583.
\bibAnnoteFile{Russ05}

\bibitem{Socolich05}
Socolich M, Lockless S, Russ W, Lee H, H GK, et~al. (2005) Evolutionary
  information for specifying a protein fold.
\newblock Nature 437:512--518.
\bibAnnoteFile{Socolich05}

\bibitem{Oates87}
Oates JF (1987) Food distribution and foraging behavior.
\newblock In: Smuts B, Cheney DL, Seyfarth RM, Wrangham RW, Struhsaker TT,
  editors, Primate societies, Chicago: University of Chicago Press. pp.
  197--209.
\bibAnnoteFile{Oates87}

\bibitem{Wrangham87}
Wrangham RW (1987) Evolution of social structure.
\newblock In: Smuts B, Cheney DL, Seyfarth RM, Wrangham RW, Struhsaker TT,
  editors, Primate societies, Chicago: University of Chicago Press. pp.
  282--298.
\bibAnnoteFile{Wrangham87}

\bibitem{Eisenberg72}
Eisenberg JF, Muckenhirn NA, Rundran R (1972) The relation between ecology a
  social structure in primates.
\newblock Science 176:863--874.
\bibAnnoteFile{Eisenberg72}

\bibitem{Schneidman05}
Schneidman E, Berry MJn, Segev R, Bialek W (2006) Weak pairwise correlations
  imply strongly correlated network states in a neural population.
\newblock Nature 440:1007--1012.
\bibAnnoteFile{Schneidman05}

\bibitem{Shlens06}
Shlens J, Field G, Gauthier JL, Grivich MI, Petrusca D, et~al. (2006) The
  structure of multi-neuron firing patterns in primate retina.
\newblock J Neurosci 26:8254--8366.
\bibAnnoteFile{Shlens06}

\bibitem{Tang08}
Tang A, Jackson D, Hobbs J, Chen W, Smith JL, et~al. (2008) A maximum entropy
  model applied to spatial and temporal correlations from cortical networks in
  vitro.
\newblock J Neurosci 28:505--518.
\bibAnnoteFile{Tang08}

\bibitem{Bethge07}
Bethge M, Berens P (2008) Near-maximum entropy models for binary neural
  representations of natural images.
\newblock In: Platt J, Koller D, Singer Y, Roweis S, editors, Advances in
  Neural Information Processing Systems 20, Cambridge, MA: MIT Press. pp.
  97--104.
\bibAnnoteFile{Bethge07}

\bibitem{Yu08}
Yu S, Huang D, Singer W, Nikolic D (2008) A small world of neuronal synchrony.
\newblock Cereb Cortex :Advance online release, doi:10.1093/cercor/bhn047.
\bibAnnoteFile{Yu08}

\bibitem{Kullback51}
Kullback S, Leibler RA (1951) On information and sufficiency.
\newblock Annals of Mathematical Statistics 22:79--86.
\bibAnnoteFile{Kullback51}

\bibitem{Friedman01}
Friedman N, Mosenzon O, Slonim N, Tishby N (2001) Multivariate information
  bottleneck.
\newblock In: Proc. ofUncertainty in Artificial Intelligence (UAI-17), San
  Mateo, CA: Morgan Kaufmann Publishers. pp. 152--161.
\bibAnnoteFile{Friedman01}

\bibitem{Slonim06}
Slonim N, Friedman N, Tishby N (2006) Multivariate information bottleneck.
\newblock Neural Comp 18:1739--1789.
\bibAnnoteFile{Slonim06}

\bibitem{Shannon49}
Shannon C, Weaver W (1949) The mathematical theory of communication.
\newblock Urbana, Illinois: University of Illinois Press.
\bibAnnoteFile{Shannon49}

\bibitem{Cover91}
Cover TM, Thomas JA (1991) Elements of information theory.
\newblock New York, NY: John Wiley \& Sons.
\bibAnnoteFile{Cover91}

\bibitem{Amari08}
Amari S (2008) Measures of correlations under changing firing rates.
\newblock Neural Comp In press.
\bibAnnoteFile{Amari08}

\bibitem{Dill85}
Dill KA (1985) Theory for the folding and stability of globular proteins.
\newblock Biochemistry 24:1501--1509.
\bibAnnoteFile{Dill85}

\bibitem{Lockless99}
Lockless SW, Ranganathan R (1999) Evolutionarily conserved pathways of
  energetic connectivity in protein families.
\newblock Science 286:295--299.
\bibAnnoteFile{Lockless99}

\bibitem{Vargas-Madraz94}
Vargas-Madrazo E, Lara-Ochoa F, Jimenez-Montano M (1994) A skewed distribution
  of amino acids at recognition sites of the hypervariable region of
  immunoglobulins.
\newblock J Mol Evol 38:100--104.
\bibAnnoteFile{Vargas-Madraz94}

\bibitem{Sarmonov62}
Sarmanov OV (1962) Maximum correlation coefficient (nonsymmetric case).
\newblock In: Selected Translations in Mathematical Statistics and Probability,
  Amer. Math. Soc., volume~2. pp. 207--210.
\bibAnnoteFile{Sarmonov62}

\bibitem{Sarmonov63}
Sarmanov OV (1963) Maximum correlation coefficient (nonsymmetric case).
\newblock In: Selected Translations in Mathematical Statistics and Probability,
  Amer. Math. Soc., volume~4. pp. 271--275.
\bibAnnoteFile{Sarmonov63}

\bibitem{Lancaster58}
Lancaster HO (1958) The structure of bivariate distributions.
\newblock Ann Math Statistics 29:719–736.
\bibAnnoteFile{Lancaster58}

\bibitem{Lancaster63}
Lancaster HO (1963) Correlation and complete dependence of random variables.
\newblock Ann Math Statistics 34:1315--1321.
\bibAnnoteFile{Lancaster63}

\bibitem{Bahadur61}
Bahadur RR (1961) A representation of the joint distribution of responses to n
  dichotomous items.
\newblock In: Solomon H, editor, Studies in Item Analysis and Prediction,
  Stanford University Press. p. 158–168.
\bibAnnoteFile{Bahadur61}

\end{thebibliography}

\end{document}